\newcount\pageformat\pageformat=2  
\newcount\pssize

\ifnum\pageformat=2
 \documentstyle[aps,prb,twocolumn,psfig]{revtex}
\else
 \documentstyle[aps,prb,preprint,psfig]{revtex}
\fi

\def\bn{ {\bf n} }

\begin{document}
\draft
\title{
Relaxation schemes for normal modes of magnetic vortices: 
Application to S-matrix
}
\author{
G.\ M.\ Wysin
}
\address{
Department of Physics, Kansas State University,
Manhattan, Kansas 66506-2601
}
\date{August 16, 2000}
\maketitle
\ifnum\pageformat=2
 \widetext
 \vskip-0.4in\hskip0.75in
 \parbox[t]{5.5in}{
\else
 \begin{abstract}
\fi
Relaxation schemes for finding normal modes of nonlinear excitations are 
described, and applied to the vortex-spinwave scattering problem in classical 
two-dimensional easy-plane Heisenberg models.  
The schemes employ the square of an effective Hamiltonian to ensure positive 
eigenvalues, together with an evolution in time or a self-consistent 
Gauss-Seidel iteration producing diffusive relaxation.
We find some of the lowest frequency spinwave modes in a circular system 
with a single vortex present.
The method is used to describe the vortex-spinwave scattering (S-matrix
and phase shifts) and other dynamical properties of vortices in ferromagnets 
and antiferromagnets, for systems larger than that solvable by 
diagonalization methods.
The lowest frequency modes associated with translation of the vortex center
are used to estimate the vortex mass. 

\ifnum\pageformat=2
 \vskip0.07in
 \noindent
 PACS numbers: 75.10.Hk; 75.40.Cx }
 \narrowtext
\else
 \end{abstract}
 \pacs{ PACS numbers: 75.10.Hk; 75.40.Cx }
\fi

\section{Introduction: Vortex Normal Modes}
The spectrum of small-amplitude vibrations in the presence of a nonlinear
excitation, such as a vortex, soliton, or other inhomogeneous magnetization,
contains information about the scattering properties, the translational 
properties, and the internal modes of vibration and instabilities of 
that object.
Any modes that acquire zero frequency, for example, as a parameter
is changed, signal a change of symmetry or phase transition, 
or perhaps a process such as magnetization reversal.
Other modes may be coupled to the translation of the center
of the object:  such translation modes contain information
about the effective mass of the excitation.
If the modes are calculated by diagonalization methods, the size
of the matrix and cpu time necessary to solve it very easily 
exceed the practical limits of any computer even for fairly
small systems.
Thus it is interesting to consider other methods for the calculation
of at least the lowest frequency modes in the presence of an
inhomogeneous state.

In particular, classical magnetic models are known to support localized
or at least partially localized nonlinear excitations, one of the
most well-known examples being the vortices of two-dimensional (2D) 
models with easy-plane (XY) anisotropy:
\begin{eqnarray} \label{Hspin}
H = & - & \frac{J}{2} \sum_{\bf n, a} (S^{x}_{\bf n} S^{x}_{\bf n+a}
+ S^{y}_{\bf n} S^{y}_{\bf n+a}
+ \lambda S^{z}_{\bf n} S^{z}_{\bf n+a}) \nonumber \\
& - & \vec{h} \cdot \sum_{\bf n} \vec{S}_{\bf n},
\end{eqnarray}
where subscripts $\bf n$ and $\bf a$ label the lattice sites 
and displacements to the nearest neighbors, and
$0 \le \lambda < 1$ determines the anisotropy, 
and we have included an applied magnetic field $\vec{h}$.
Quite generally, linearization of 
\ifnum\pageformat=2
  \vbox to 140pt{\noindent}
\fi

\noindent
the associated equations of motion about
a nonlinear excitation such as a vortex will lead to a normal mode or 
spinwave problem, with an operator $\cal M$ producing the time evolution of 
a spinwave wavefunction $\Psi$:
\begin{equation} \label{Schrod0}
{d \over dt} \Psi = {\cal M} \Psi.  
\end{equation}
The calculation of the spectrum of spinwaves in the presence
of a vortex by diagonalization was accomplished\cite{WysinVolkel95} 
for small circular systems up to a radius $R\approx 20a$, where
$a$ is the lattice constant.
The method was used to determine vortices' spinwave spectrum for  
ferromagnets (FM, $J>0$) and antiferromagnets (AFM, $J<0$)
on a lattice.\cite{WysinVolkel96} 
In addition to usual scattering states,
it was found that FM and AFM vortices possess an internal mode whose
frequency goes to zero at a critical anisotropy value $\lambda_c$,
which is lattice dependent ($\approx 0.7034$ for square lattice).
The mode is associated with the instability of out-of-plane
vortices (those with $S^z\ne 0$) to change to in-plane vortices 
$S^z=0$ everywhere) at strong easy-plane anisotropy 
($\lambda<\lambda_c$).\cite{Wysin+88,vortex-lat,Wysin94}
Other modes found, which tend to have high intensity at the vortex core, 
are associated with the translation of the vortex center of mass,
and can be related to a collective coordinate description for
vortex dynamics.\cite{Mertens+97,Ivanov+98}

Analytic calculations of the vortex normal modes, using a continuum limit, 
have partially described the scattering spectrum.\cite{Costa92,Pereira93,Pereira+96,PP96}
In the earliest calculations, the phase shifts of spinwaves scattered by 
FM in-plane vortices were calculated, using a Born approximation, 
for the exchange anisotropy model (\ref{Hspin}) 
considered here \cite{Costa92} and
for a similar FM model with site-anisotropy.\cite{Pereira93}
The weak point of these calculations is that they rely on the choice
of a short distance cutoff for the Born approximation integrals.
For the AFM in-plane vortices, Pereira {\it et al.}\cite{Pereira+96}
were able to make an exact calculation of the scattering phase shifts,
for the optical spinwave branch, without the need for a cutoff.
In the continuum limit equations, AFM vortices do not scatter the acoustic
branch spinwaves; the associated phase shifts vanish.

The validity of the continuum limit and other approximations, however,
can only be tested by comparison with numerical calculations.
For example, Ivanov {\it et al.} \cite{Ivanov+96} found by analytic means 
that the vortices of the continuum AFM model for $\lambda$ near 1
should possess a true internal local mode.
It was shown only through numerical diagonalization that this mode
is identical to the vortex instability mode mentioned above.
Ivanov {\it et al.} were also able to calculate the scattering (S-matrix)
by a numerical shooting solution for the continuum theory equations of motion.
The S-matrix was not calculated from the numerical diagonalization data,
however, because the system size that could be solved was too small.
%

For $\lambda_c < \lambda \approx 1$, 
the nonzero $S^z$ component of the out-of-plane 
vortex leads to a smooth structure decaying over the length scale 
\begin{equation}
\label{rv}
r_v = \sqrt{\lambda\over 1-\lambda},
\end{equation}
with no singularity at the vortex center.
Thus, the continuum description used in Ref.\ \onlinecite{Ivanov+96}
is quite accurate. 
At the other limit, $\lambda < \lambda_c$, 
the in-plane vortex spin field is singular at the vortex center.
In this case, the continuum limit cannot
describe very well the spin dynamical motions near the vortex core,
without the introduction of some kind of lower radius cutoff.
The length scale in the problem approaches zero, and derivatives of 
the static vortex spin structure on a lattice are not small in 
the core region, violating the usual continuum limit assumptions
about small spatial derivatives across one lattice constant.
Nevertheless, even for the in-plane AFM vortices,  
a true local mode of the vortex is present,\cite{Wysin+98}
and can be approximately described in continuum theory, but 
only if a cutoff is introduced. 
Therefore, for both FM and AFM {\em in-plane} vortices, it is
interesting to consider the calculation of the S-matrix
(or equivalently, phase shifts) for the lattice 
system, and compare with continuum scattering theory. 
%

It has been realized\cite{Wysin96} that the translational modes that come
from the spinwave spectra have unusual properties and can be used
to calculate a vortex effective mass.  
Although only a limited size of system could be studied ($R<20a$)
it was found that the FM vortex mass increases faster than $\ln R$, 
for in-plane as well as out-of-plane vortices.   
AFM in-plane vortices also were found to have a mass increasing faster 
than $\ln R$, however, it appeared that the mass of AFM out-of-plane vortices
may actually reach a finite limit for large $R$.  
It is also possible that AFM vortex mass tends to zero for $\lambda\to 1$,
the isotropic limit.
%

More recently, a newer theory for the FM out-of-plane vortex mass and 
related collective coordinate description has been developed by Mertens
et al.\cite{Mertens+97}
There, a vortex mass approximately independent of system size
was determined, both by including effects due to an image
vortex outside the studied system, and also, by using a dynamical
equation of motion higher than second order in time.
This newer theory appears to describe well the individual vortex
dynamics by assigning a mass, gyrovector, and a new higher order
gyro-tensor to the vortex, considered as a particle-like object.
A careful application of the theory was made by Ivanov et 
al,\cite{Ivanov+98} using spinwave translation mode data
obtained from another numerical scheme for finding some
of the lowest modes.
In Ref. \onlinecite{Ivanov+98}, Wielandt's version of inverse
iteration procedure was used:  an operator $({\cal M}-\omega I)^{-1}$
was applied to an initial randomly chosen vector, which after
many iterations converges to an eigenvector of ${\cal M}$, provided
the frequency $\omega$ is updated appropriately.  
It is also a type of relaxation procedure similar to what we 
describe here, with great speed and memory advantages over the
diagonalization method.
This vortex mass theory, however, does not apply directly to the 
AFM vortices, nor to the in-plane vortices for both FM and AFM models. 
%

In this paper we present and apply some numerical methods for finding 
the normal modes of a magnetic vortex on a lattice, which are applicable 
to systems larger than those solvable by diagonalization schemes.
We use these methods to analyze the FM and AFM spinwave-vortex scattering 
problems for in-plane vortices.
In addition to the calculation of the S-matrix and phase shifts, we present 
results on the translational modes and make estimates of in-plane
vortex mass.
Analysis of out-of-plane vortices will be considered 
elsewhere.\cite{BorisFuture}
%

The paper is organized as follows.
First, in Sec.\ \ref{Brief}, a short overview of relaxation schemes is given.
In Sec.\ \ref{Spinwaves}, we describe the magnetic spinwave equations 
of motion in the presence of a {\em nonuniformly} magnetized configuration.
This is followed in Sec.\ \ref{Methods} by a description of the 
relaxation schemes using time evolution and Gauss-Seidel iteration.
In Sec.\ \ref{Spectra} the schemes are used to calculate the 
vortex-spinwave S-matrix, for a single vortex in a circular system. 
After a summary of the translational dynamics of magnetic vortices,
in Sec.\ \ref{Mass} we present and analyze the results obtained for 
the FM and AFM in-plane vortex mass.
%

\section{Relaxation Methods for Eigenfunctions}
\label{Brief}
The basic idea of these relaxation schemes is the following:
The eigenfrequencies of a linear spinwave problem, such as 
Eq.\ (\ref{Schrod0}), always come in positive/negative pairs, 
corresponding essentially to creation and annihilation operators, 
respectively.
Thus, its time evolution leads to oscillatory behavior, starting from
an arbitrary initial condition.
It is interesting, however, to consider how the evolution can be made 
diffusive, such that high ({\em absolute magnitude}) frequency modes 
decay away faster than low frequency modes.
For example, a Schr\"odinger equation with the replacement 
$it\rightarrow\tau$, becomes a diffusion equation in imaginary 
time $\tau$ (because it is a parabolic equation).
It is well known that its imaginary time evolution will lead to the 
lowest frequency wavefunction, i.e., the ground state of the 
associated Hamiltonian. 
In the equations of motion (\ref{Schrod0}), such a replacement 
$it\rightarrow\tau$, unfortunately, does not lead to a 
diffusion-like equation, due to the presence of the two fundamental 
solutions (creation/annihilation operators) behaving as 
$e^{\pm i\omega_k t}\rightarrow e^{\pm \omega_k \tau}$ 
(because the spinwave equations are essentially hyperbolic).
There is a fundamental solution growing with $\tau$ for every mode $k$,
leading to numerical instability in imaginary time.  
However, if one uses the {\em square} of the spinwave Hamiltonian to 
evolve forward in real time, then both fundamental solutions associated 
with mode $k$ decay as $e^{- \omega_k^2 t}$.
Evolution forward in time will lead to the lowest frequency eigenmode.
By combining with Gram-Schmidt orthogonalization, it is then possible to 
iterate this procedure and generate a small set of the lowest frequency 
eigenmodes.  
The method offers a drastic savings in memory needed compared to 
numerical diagonalization, at the expense of rather long cpu time 
due to its diffusive behavior.  
Also note, it is not essential that the evolution be performed in time.
We have also found that the eigenmodes can be found even more quickly 
by using a Gauss-Seidel iteration of the spinwave equations, where the 
unknown frequency $\omega$ within those equations is determined 
self-consistently during the iteration.

\section{2D Easy-Plane Magnetic Model and Spinwave Equations}
\label{Spinwaves}
We review briefly the derivation of the linearized spinwave problem
for normal modes on a vortex, or other nonuniform state.
The nonlinear equations of motion from (\ref{Hspin}) are simple:
\begin{equation} \label{SpinMotion}
\dot{\bf S}_{\bf n}= {\bf S}_{\bf n} \times 
\left[ \vec{h} + 
\sum_{\bf n'=n+a} \tilde{J}\cdot{\bf S}_{\bf n'} \right],
\end{equation}
where $\tilde{J}$ is a constant diagonal exchange matrix:
\begin{equation} \label{Jmatrix}
\tilde{J} = J 
\left( \begin{array}{ccc}
           1   &  0  &  0     \\
           0   &  1  &  0     \\
           0   &  0  &  \lambda  
\end{array}  \right)
\end{equation}
and the spins are considered as column vectors.
A vortex is a particular static configuration $\{ {\bf S}_{\bf n}^0 \}$ that 
solves this equation with a zero time derivative.  
The structure of in-plane and out-of-plane vortices has
been described elsewhere using continuum theory\cite{vortex-th} and 
in the presence of a lattice.\cite{vortex-lat,Wysin94}
Because the spin length is a conserved quantity, there are 
effectively only two degrees of freedom per spin, and it is usually 
convenient to describe the static structure using an in-plane angle 
$\phi_{\bf n}^0$ and an out-of-plane angle $\theta_{\bf n}^0$, in
planar spherical coordinates:
\begin{equation} \label{spherical}
{\bf S}_{\bf n}^0 = S \left(
  \begin{array}{c} 	\cos\theta_{\bf n}^0 \cos\phi_{\bf n}^0 \\
			\cos\theta_{\bf n}^0 \sin\phi_{\bf n}^0 \\
			\sin\theta_{\bf n}^0 
  \end{array}		\right) .
\end{equation}
Then one could obtain the spinwave effective equations of motion by 
assuming a perturbation to this structure in the form,
\begin{mathletters} \label{pert1}
\begin{equation} 
 \phi_{\bf n} = \phi_{\bf n}^0 + \varphi_{\bf n},
\end{equation}
\begin{equation}
 \theta_{\bf n} = \theta_{\bf n}^0 + \vartheta_{\bf n},
\end{equation}
\end{mathletters}
where the equations of motion are now linearized in terms of 
$\varphi_{\bf n}$ and $\vartheta_{\bf n}$.
%

In Ref. \onlinecite{WysinVolkel95} a slightly different but equivalent 
description was used, which we follow here.
The unperturbed spins of the static vortex structure are considered to 
define {\em local} quantization axes $\tilde{z}_{\bf n}$, specifically, 
\begin{equation} \label{Szero}
{\bf S}_{\bf n}^0 = S \tilde{z}_{\bf n}. 
\end{equation}
Then the perturbation of this structure involves fluctuations orthogonal 
to the $\tilde{z}_{\bf n}$-axis, along two new local axes 
$\tilde{x}_{\bf n}$ and $\tilde{y}_{\bf n}$.  
The $\tilde{x}_{\bf n}$-axis is taken to be along the direction defined by
the cross product  $\tilde{x}_{\bf n}=z_{\bf n} \times \tilde{z}_{\bf n}$, 
which is within the original xy (easy) plane.  
Quantities $x_{\bf n}$, $y_{\bf n}$, and $z_{\bf n}$ here refer to the
original coordinates of Hamiltonian (\ref{Hspin}).
Then $\tilde{y}_{\bf n}=\tilde{z}_{\bf n} \times \tilde{x}_{\bf n}$
makes the last member of the local orthogonal set.
The perturbation of the static vortex structure can then be expressed in 
terms of its spin components along these new local axes:
\begin{equation} \label{perturb}
{\bf S}_{\bf n} = S_{\bf n}^{\tilde{z}} ~ \tilde{z}_{\bf n} 
                + S_{\bf n}^{\tilde{x}} ~ \tilde{x}_{\bf n}
	        + S_{\bf n}^{\tilde{y}} ~ \tilde{y}_{\bf n}  . 
\end{equation}
A short calculation shows that these are related to the angular 
perturbation coordinates by
\begin{equation}
S_{\bf n}^{\tilde{x}} = S \varphi_{\bf n} \cos \theta_{\bf n}^{0}, ~~~
S_{\bf n}^{\tilde{y}} = S \vartheta_{\bf n}.
\label{varangles} \end{equation}
The variables $S_{\bf n}^{\tilde{x}}$ and $\varphi_{\bf n}$ relate to purely
in-plane spin motions, while $S_{\bf n}^{\tilde{y}}$ and $\vartheta_{\bf n}$
measure the out-of-easy-plane tilting, relative to the unperturbed vortex.  

In Refs. \onlinecite{WysinVolkel95} and \onlinecite{WysinVolkel96} it is shown 
that under the assumptions $S_{\bf n}^{\tilde{x}} \ll 1$,  
$S_{\bf n}^{\tilde{y}} \ll 1$, $S_{\bf n}^{\tilde{z}} \approx S$, this 
notation leads to a usual linear time evolution problem, which we write 
here in the form:
\begin{equation} \label{Sdot}
{d \over dt} 
\left( 
\begin{array}{c} S^{\tilde{x}}_{\bf n} \\ S^{\tilde{y}}_{\bf n} \end{array} 
\right) =
\sum_{\bf m=n,n' } \left( 
\begin{array}{cc}  M^{\tilde{x}\tilde{x}}_{\bf n m } 
                 & M^{\tilde{x}\tilde{y}}_{\bf n m } \\ 
                   M^{\tilde{y}\tilde{x}}_{\bf n m } 
                 & M^{\tilde{y}\tilde{y}}_{\bf n m }  \end{array}
\right)  \left(
\begin{array}{c} S^{\tilde{x}}_{\bf m} \\ S^{\tilde{y}}_{\bf m} \end{array}
\right)
\end{equation}
or equivalently in components,
\begin{equation} \label{Sdotcomp}
\dot{S}_{\bf n}^{\alpha} = 
\sum_{\bf m=n,n'} \sum_{\beta}
                M_{\bf n m}^{\alpha\beta} S_{\bf m}^{\beta}
\end{equation}
where $\alpha$ and $\beta$  range over the set $\{ \tilde{x}, \tilde{y} \}$, 
and the site ${\bf m}$ must either be equal to site ${\bf n}$ or one of 
its neighbors, ${\bf n'=n+a}$. 
The matrix elements $M_{\bf n m}^{\alpha\beta}$ are: 
\begin{mathletters}\label{elements}
\begin{eqnarray} 
& M&_{\bf n n}^{\tilde{x}\tilde{x}} = M_{\bf n n}^{\tilde{y}\tilde{y}} = 0,\\ 
& M&_{\bf n n}^{\tilde{x}\tilde{y}}  = -M_{\bf n n}^{\tilde{y}\tilde{x}} =  
 (h_x \cos\phi_{\bf n}^0 + h_y \sin\phi_{\bf n}^0 ) p_{\bf n}^0 
+ h_z m_{\bf n}^0 \nonumber \\
&+& JS \sum_{\bf n'=n+a} 
        \left[ 
                p_{\bf n}^0 p_{\bf n'}^0 \cos(\phi_{\bf n}^0-\phi_{\bf n'}^0)
             +  \lambda m_{\bf n}^0 m_{\bf n'}^0
        \right],  \\
&M&_{\bf n, n'}^{\tilde{x}\tilde{x}}  = 
              JS ~ m_{\bf n}^0 \sin(\phi_{\bf n}^0-\phi_{\bf n'}^0), \\
&M&_{\bf n, n'}^{\tilde{y}\tilde{y}}  = 
              JS ~ m_{\bf n'}^0 \sin(\phi_{\bf n}^0-\phi_{\bf n'}^0), \\
&M&_{\bf n, n'}^{\tilde{x}\tilde{y}}  = 
     - JS \left[ m_{\bf n}^0 m_{\bf n'}^0 \cos(\phi_{\bf n}^0-\phi_{\bf n'}^0)
        + \lambda p_{\bf n}^0 p_{\bf n'}^0 
          \right]  \\
&M&_{\bf n, n'}^{\tilde{y}\tilde{x}}  =  
        JS \cos(\phi_{\bf n}^0-\phi_{\bf n'}^0) .
\end{eqnarray}
\end{mathletters}
For simplification, we use the definitions of in-plane projection and 
out-of-plane magnetization,
\begin{mathletters}
\begin{eqnarray}
p_{\bf n}^0 & \equiv & \sqrt{1-(S_{\bf n}^{z 0}/S)^2} = \cos\theta_{\bf n}^0, \\
m_{\bf n}^0 & \equiv & S_{\bf n}^{z 0}/S = \sin\theta_{\bf n}^0 .
\end{eqnarray}
\end{mathletters}
%

The operator on the right hand side of the linear equations of motion,
Eq.\ (\ref{Sdotcomp}), is not Hermitian.  (For example, consider the case 
$\lambda=0$, with all $\theta_{\bf n}^0=0$.)  
This led to complications in the previous numerical diagonalization 
calculations.  
Its eigenvalues are pure imaginary, in complex conjugate pairs; this results
in the usual spinwave solutions oscillatory in time.  
However, this implies that its square (which need not be Hermitian)
has pure real eigenvalues, a fact we will exploit below in Sec.\ \ref{Methods}.  

In Refs. \onlinecite{WysinVolkel95} and \onlinecite{WysinVolkel96}, 
the variables $S^{\tilde{x}}$ and $S^{\tilde{y}}$ were considered as
quantum operators, in a semiclassical picture.
Then an eigenmode labeled by index $k$, has a creation operator 
$B^{\dagger}_k$ that is
an appropriate linear combination, with complex expansion coefficients, 
$w_{k, \bf n}^{\it 1}, w_{k, \bf n}^{\it 2}$,
\begin{equation}\label{Bdagger}
B^{\dagger}_{k} = \sum_{\bf n} \left[ 
   w_{k, \bf n}^{\it 1} S_{\bf n}^{\tilde{x}} 
+  w_{k, \bf n}^{\it 2} S_{\bf n}^{\tilde{y}} \right],
\end{equation}
such that its time evolution is simple, with eigenfrequency $\omega_k$:
\begin{equation}
\dot{B}^{\dagger}_k = i \omega_k B^{\dagger}_k .
\end{equation}
The corresponding annihilation operator is $B_k=(B_k^{\dagger})^{*}$,
since the spin operators are real.
The local spin fluctuations are given from the inverse relationship,
\begin{mathletters}
\label{SxSy}
\begin{equation}
S_{\bf n}^{\tilde{x}} = i\hbar S \sum_k 
\left( w_{k, \bf n}^{\it 2} B_k - w_{k, \bf n}^{\it 2 \; *} B_k^{\dagger} \right),
\end{equation}
\begin{equation}
S_{\bf n}^{\tilde{y}} = -i\hbar S \sum_k 
\left( w_{k, \bf n}^{\it 1} B_k - w_{k, \bf n}^{\it 1 \; *} B_k^{\dagger} \right),
\end{equation}
\end{mathletters}
Note that the italic superscripts on the 
$w_{\bf n}^{\alpha}, ~ \alpha={\it 1, 2}$ in this article should 
not be confused with powers, because this is a linear problem.

The collection of coefficients describes the wavefunction, and satisfies
\begin{equation} \label{w1w2}
i \omega_k
\left( 
\begin{array}{c} w^{\it 1}_{k, \bf n} \\ w^{\it 2}_{k, \bf n} \end{array} 
\right) =
\sum_{\bf m=n,n' } \left( 
\begin{array}{cc}  M^{\tilde{x}\tilde{x}}_{\bf n m } 
                 & M^{\tilde{y}\tilde{x}}_{\bf n m } \\ 
                   M^{\tilde{x}\tilde{y}}_{\bf n m } 
                 & M^{\tilde{y}\tilde{y}}_{\bf n m }  \end{array}
\right)  \left(
\begin{array}{c} w^{\it 1}_{k, \bf m} \\ w^{\it 2}_{k, \bf m} \end{array}
\right)
\end{equation}
The matrix on the RHS is the transpose of that in Eq.\ (\ref{Sdot}).
Thus the two systems clearly have the same eigenvalues, however, we prefer
to use the $w^{\it 1}, w^{\it 2}$ notation, together with its underlying 
creation and annihilation operators, because the overall wavefunction 
orthogonalizations are determined by related commutators.\cite{orthoNote}
A wavefunction can be considered as a column vector of these coefficients:
\begin{equation} \label{wavef}
\Psi_k = \left( \begin{array}{c} \psi_{k, \bf 1} \\ \psi_{k, \bf 2} \\ . \\ 
                                  . \\ \psi_{k, \bf N} 
              \end{array}                       \right),
\end{equation}
where some labeling from 1 to $N$ has been given to the $N$ sites, each 
of which has a local two-component wavefunction holding the $S^{\tilde{y}}$
($w^{\it 1}$) and $S^{\tilde{x}}$ ($w^{\it 2}$) fluctuations:
\begin{equation}
\psi_{k, \bf n} = \left( 
          \begin{array}{c} w_{k, \bf n}^{\it 1} \\  w_{k, \bf n}^{\it 2} \end{array} 
               \right).
\end{equation}
The dynamics of these individual wavefunctions can be written in terms of
the matrices appearing in Eq.\ (\ref{w1w2}), which we denote here for
convenience as ${\cal M}_{\bf n m}$:
\begin{equation}
i \omega_k  ~ \psi_{k, \bf n} = \sum_{\bf m=n,n'} 
                                {\cal M}_{\bf n m} ~ \psi_{k, \bf m}.
\end{equation}
Then we can see that there exists an operator $\cal M$ composed of the 
individual ${\cal M}_{\bf n m}$ matrices, with real elements, that evolves 
the entire wavefunction:
\begin{equation}\label{Schrod}
i \omega_k  \Psi_k = {\cal M}  \Psi_k .
\end{equation}
This is the linear problem to be solved.

We have previously solved this non-Hermitian problem, Eq.\ (\ref{Schrod}), 
through numerical diagonalization, using the EISPACK routine RG() for 
diagonalizing real-general matrices (no symmetry assumed).
In this way the full spectrum could be found, however, the matrix $\cal M$
is sparse but of size $2N \times 2N$, where $N$ is the number of lattice sites.
A circular square lattice system with radius $R=20a$ has $N=1264$.
This leads to a matrix with 6390784 elements, which when stored in double
precision (8 bytes per number), requires 49 MB RAM.  
In practice, it is also necessary to store another array of equal size that 
contains the calculated eigenmodes, thus the required storage for a 
calculation at one value of $\lambda$ is twice this number.
This gets incremented by another 49 MB if one wants to compare eigenmodes at
nearby values of $\lambda$ for tracking their changes with anisotropy.
For these reasons, on a machine with 128 MB RAM, the upper limit for this
calculation was system radius $R=$19--20, depending on the desired results.

%
Next, we consider how the eigenmodes can be found through squaring the 
operator of the RHS of Eq.\ (\ref{Schrod}), combined with introducing a 
fictitious time evolution, resulting in considerable memory savings.

\section{Relaxation Method} 
\label{Methods}

\subsection{Time Evolution}

A mode $\Psi_k$ that solves Eq.\ (\ref{Schrod}) is a fundamental solution of
\begin{equation} \label{Psidot}
{d \over dt} \Psi = {\cal M} ~ \Psi .
\end{equation}
with time dependence $e^{i\omega_k t}$, determined by the corresponding 
eigenvalue of ${\cal M}$, $i\omega_k$. 
Applying the ${\bf \cal M}$ operation again leads to an equivalent equation,
\begin{equation} \label{Psidotdot}
{d^2 \over dt^2} \Psi = {\cal M}^2 ~ \Psi . 
\end{equation}
Both of these equations have fundamental oscillatory solutions varying as
$\Psi_k(t) =\Psi_k(0) e^{i \omega_k t}$, i.e., the usual spinwaves.
The advantage is that operator ${\cal M}^2$ has purely {\em real, negative}
eigenvalues, in contrast to the purely imaginary eigenvalues (both positive
and negative imaginary parts) of operator ${\cal M}$.

Now consider that we invent a fictitious time evolution, by rather 
arbitrarily changing the LHS of Eq.\ (\ref{Psidotdot}) to a first 
time derivative:
\begin{equation}\label{newtime}
{d \over dt} \Psi = {\cal M}^2 ~ \Psi .
\end{equation}
The fundamental solutions here are determined by exponentials involving 
the eigenvalues of ${\cal M}^2$, which are $-\omega_k^2$.  
Thus, the general solution from an arbitrary initial state is
\begin{equation}\label{PsiSol}
\Psi(t) = \sum_{k} 
\left[ c_k^{+} \Psi_k^{+} + c_k^{-} \Psi_k^{-} \right] 
e^{-\omega_k^2 t} ,
\end{equation}
where the coefficients $c_k^{\pm}$ are determined by the initial state, 
and the $\Psi_k^{\pm}$ are normalized eigenfunctions of ${\cal M}$, 
with respective eigenvalues $\pm i \omega_k$, corresponding to the 
creation and annihilation operators for mode $k$.
(Of course, $\Psi_k^{\pm}$ have the same eigenvalues, $-\omega_k^2$,  
with respect to the operator ${\cal M}^2$.)
All modes decay away in time, leaving the lowest frequency modes present 
in the initial state to dominate at large time.
In practice, the time evolution will be started from a randomly 
chosen initial state (i.e., randomly assigned values of the set of 
$\{w_{\bf n}^{\it 1}, w_{\bf n}^{\it 2} \}$ coefficients), to ensure 
a nonzero overlap with the lowest frequency mode.
The total wavefunction $\Psi(t)$ will be re-normalized to unity 
periodically to avoid numerical underflow, by enforcing the overlap 
appropriate for a non-Hermitian operator described below 
(Sec.\ \ref{GramSchmidt}).
\begin{equation}\label{HermOver}
\left< \Psi | \Psi \right> = 1.
\end{equation}
%
%
We have used a second order Runge-Kutta time integration scheme to
solve Eq.\ (\ref{Psidotdot}), and found that a time step of $0.02/JS$  
is adequate to provide stability and insure convergence to an eigenstate.

At large time, the system will be relaxed to a linear combination of 
creation and annihilation wavefunctions:
\begin{equation}\label{combo}
\Psi = c_{k_0}^{+} \Psi_{k_0}^{+} + c_{k_0}^{-} \Psi_{k_0}^{-},
\end{equation}
where $k_0$ labels the lowest mode that was present in the intial state.
The frequency for this combination is evaluated in the usual way,
\begin{equation}\label{freq}
{\omega_{k_0}}^2 =  -  { \left< \Psi | {\cal M}^2 | \Psi \right> 
               \over \left< \Psi | \Psi \right> }.
\end{equation}
The creation and annihilation wavefunctions can be separated by an 
application of operator ${\cal M}$:
\begin{equation}\label{Mcombo}
{\cal M}\Psi = i \omega_{k_0} 
 \left( c_{k_0}^{+} \Psi_{k_0}^{+} - c_{k_0}^{-} \Psi_{k_0}^{-} \right),
\end{equation}
Combining the two equations (\ref{combo}) and (\ref{Mcombo}) leads to 
\begin{equation}\label{separate}
c_{k_0}^{\pm} \Psi_{k_0}^{\pm} = { {i\omega_{k_0} \Psi \pm {\cal M}\Psi} 
                                   \over 2 i \omega_{k_0}  }.
\end{equation}
This finally will be followed by a normalization appropriate to the 
non-Hermitian operator ${\cal M}$, to be described below
[Eq.\ (\ref{overlap})].

\subsection{Orthogonality of Modes and Gram-Schmidt Process} 
\label{GramSchmidt}
To obtain higher modes, it is necessary to evolve forward in time 
together with a Gram-Schmidt orthogonalization to any previously 
found lower frequency modes, denoted here by index $\ell$.   
We have found that the necessary way to do this is by making the current
wavefunction $\Psi(t)$ orthogonal to {\sl both} $\Psi_{\ell}^{+}$ and
$\Psi_{\ell}^{-}$, rather than to a somewhat arbitrary combination such 
as in Eq.\ (\ref{combo}).
Equivalently, both constants $c_{\ell}^{+}$ and $c_{\ell}^{-}$ must be 
enforced to be zero in the current wavefunction, for all modes $\ell$ 
already determined.
To obtain these {\em two} conditions, it is necessary to use the 
overlap appropriate to the non-Hermitian operator ${\cal M}$, 
since $\Psi_{\ell}^{+}$ and $\Psi_{\ell}^{-}$ are distinct 
eigenfunctions of ${\cal M}$ corresponding to its distinct 
eigenvalues $\pm i \omega_{\ell}$.  
%

The overlap of two eigenmodes of ${\cal M}$ can be defined in one of 
two equivalent ways.
The first is to realize that an annihilation ($B_j$) and creation 
($B_k^{\dagger}$) operator, with associated wavefunctions
$\Psi_j^{-}$ and $\Psi_k^{+}$, must have a commutator,
\begin{equation}\label{commute}
[ B_j , B_k^{\dagger} ] = \delta_{j,k}.
\end{equation}
From the definition, Eq.\ (\ref{Bdagger}), with the help of the basic
spin commutators $[S_{\bf n}^{\tilde{x}} , S_{\bf n'}^{\tilde{y}} ] =
i \delta_{\bf n,n'} S_{\bf n}^{\tilde{z}} \approx i \delta_{\bf n,n'} S$, 
this is equivalent to:
\begin{equation}\label{overlap}
[ B_j , B_k^{\dagger} ] = i S \sum_{\bf n} 
        \left(
   w_{j, \bf n}^{\it 1\;*} w_{k, \bf n}^{\it 2}
 - w_{j, \bf n}^{\it 2\;*} w_{k, \bf n}^{\it 1}  
        \right) 
= \left< \Psi_j | \Psi_k \right>.
\end{equation}
In general, this is the definition of the overlap used between two
arbitrary states $j$ and $k$.
%

The second way to define the overlap, equivalent to this, is to make use
of the left and right eigenvectors of ${\cal M}$.
The solutions $\Psi_k$ of Eq.\ (\ref{Schrod}), where ${\cal M}$ acts 
to the right, are the right eigenvectors.    
For each eigenvalue $i\omega_k$, there also exists a corresponding left 
eigenvector, ${ }_k \! \Psi$ satisfying
\begin{equation}\label{LSchrod}
 i\omega_k ~  { }_k \! \Psi =  { }_k \! \Psi {\cal M}.
\end{equation}
Each creation operator as well as each annihilation operator has
a distinct left eigenvector.  
For a Hermitian problem, the left eigenvectors are obtained by
the familiar relationship, ${ }_k \! \Psi = \Psi_k^{\dagger}$,
and then overlaps are in the usual form $\left< \Psi_j | \Psi_k \right>
= { }_j \! \Psi \Psi_k = \Psi_j^{\dagger} \Psi_k$.

For this problem, although ${\cal M}$ is not Hermitian, there exists a
real matrix ${\cal A = - A}^{-1}$, that transforms
${\cal M}$ to $-{\cal M}^{\dagger}$:
\begin{equation}\label{sim}
{\cal A M A}^{-1} = - {\cal M}^{\dagger} 
\end{equation}
where ${\cal A}$ is diagonal in $2\times 2$ submatrices:\cite{SignNote}
\begin{equation} \label{Adef}
{\cal A} = \left(
\begin{array}{cccc}
 \alpha  &   0      &    0    & ... \\
   0     &  \alpha  &    0    & ... \\
   0     &    0     &  \alpha & ... \\
   .     &    .     &    .    & ... \\
   .     &    .     &    .    & ... \\
\end{array}
\right),  \quad\quad  
\alpha=\left(
\begin{array}{rr} 0 & -1 \\ 
                  1 &  0 \\  \end{array} \right).
\end{equation}
Then using Eq.\ (\ref{sim}) in Eq.\ (\ref{Schrod}),  rearranging 
and comparing with Eq.\ (\ref{LSchrod}) determines a relationship 
between left and right eigenvectors,
\begin{equation}\label{leftright}
{ }_k \! \Psi = i \left( {\cal A} \Psi_k \right)^{\dagger},
\end{equation}
where the $\dagger$ operation is the usual Hermitian conjugate
(transpose of complex conjugate), and the factor of $i$ has been 
added for consistency with Eq.\ (\ref{overlap}).
Then the overlap between two states $\Psi_j$ and $\Psi_k$ is defined 
by the scalar product of a left with a right eigenvector:
\begin{equation}\label{overlap1}
\left< \Psi_j | \Psi_k \right> =
 { }_j \! \Psi  \Psi_k  = 
i \left( {\cal A} \Psi_j \right)^{\dagger} \Psi_k =
i \Psi_j^{\dagger} {\cal A}^{\dagger} \Psi_k .
\end{equation}
which is seen to give exactly the expression in (\ref{overlap}).

The application of Gram-Schmidt orthogonalization to remove the 
previously found modes (labeled by $\ell$) proceeds according to 
replacing the current $\Psi(t)$ by
\begin{equation}\label{GS}
\Psi(t) \rightarrow \Psi(t) 
- \sum_{\ell} \left[ \left< \Psi_{\ell}^{+} | \Psi(t) \right> \Psi_{\ell}^{+}
                   + \left< \Psi_{\ell}^{-} | \Psi(t) \right> \Psi_{\ell}^{-}
              \right].
\end{equation}
Combined with evolution forward in time and periodic renormalization
according to Eq.\ (\ref{HermOver}), the wavefunction will evolve to
the next low-frequency mode, of the form Eq.\ (\ref{combo}). 
Continuing to iterate the above procedures, a small set of the lowest
frequency modes (about 5-10) can be determined with double precision
accuracy in reasonable time for systems of size $100\times 100$.

\subsection{Self-Consistent Gauss-Seidel Scheme}
The eigenmodes of Eq.\ (\ref{Schrod}) alternatively can be found 
by a Gauss-Seidel type of iteration, not applied to that equation, 
but rather, to the equation with the squared operator, 
\begin{mathletters}\label{Msquared} 
\begin{equation}
{\cal H}={\cal M}^2,
\end{equation}
\begin{equation}
{\cal H} \Psi_k = -\omega_k^2 \Psi_k. 
\end{equation}
\end{mathletters}
This Gauss-Seidel scheme to be described is about twice as fast 
(in total cpu time) as the time evolution using second order Runge 
Kutta described above.
It is based on using the matrix elements of ${\cal H}$, and in particular,
one needs to single out the matrix elements that couple a site to itself
(${\cal H}_{\bf n,n}$) from the intersite couplings ($W_{\bf n}$).
We write this equation as a set of equations with the form,
\begin{mathletters}\label{M2matrix}
\begin{eqnarray}
-\omega^2 w_{\bf n}^{\it 1} & = & {\cal H}_{\bf n,n}^{\it 11} w_{\bf n}^{\it 1} +
			{\cal H}_{\bf n,n}^{\it 12} w_{\bf n}^{\it 2} + W_{\bf n}^{\it 1} \\
-\omega^2 w_{\bf n}^{\it 2} & = & {\cal H}_{\bf n,n}^{\it 21} w_{\bf n}^{\it 1} +
			{\cal H}_{\bf n,n}^{\it 22} w_{\bf n}^{\it 2} + W_{\bf n}^{\it 2}
\end{eqnarray}
\begin{eqnarray}
W_{\bf n}^{\alpha} & = & \sum_{\bf n''} \sum_{\beta=1,2} 
		{\cal H}_{\bf n,n''}^{\alpha\beta} w_{\bf n''}^{\beta} , 
\end{eqnarray}
\end{mathletters}
where the sites ${\bf n''}$ include all first and second nearest neighbors
of site ${\bf n}$, but {\em not} the site ${\bf n}$. 
The second nearest neighbor interations result directly from squaring the
nearest neighbor operator, ${\cal M}$.
These matrix elements of ${\cal H}$ are sums of quadratic products,
\begin{equation}
{\cal H}_{\bf n,n''} = \sum_{\bf m=n,n',n''} 
     {\cal M}_{\bf n,m} {\cal M}_{\bf m,n''},
\end{equation}
where the summation is over any sites ${\bf m}$ that are either first 
neighbors of sites ${\bf n}$ and ${\bf n''}$, or, these sites themselves. 
Due to the structure of the matrix elements of ${\cal M}$, 
the on-site, off-diagonal elements 
${\cal H}_{\bf n,n}^{\it 12}, \quad {\cal H}_{\bf n,n}^{\it 21}$,
are both zero, while the corresponding diagonal elements are equal:
${\cal H}_{\bf n,n}^{\it 11} = {\cal H}_{\bf n,n}^{\it 22}$.  
%
 
Now in the spirit of Gauss-Seidel iteration, we can try to ``solve'' for
the new values,  ${\psi_{\bf n}}'$, with some certain freedom of choice.
That is, the terms involving $\omega^2$ can be taken as explicit 
or implicit, depending on whether we consider them to involve the 
wavefunction at the previous step or the wavefunction at the 
current step being solved.
We have found for this problem, that the iteration is more stable 
if the $\omega^2$ terms are explicit, and the equation is re-arranged
for each component $\alpha={\it 1, 2}$ as 
\begin{equation}\label{RawIter}
{w_{\bf n}^{\alpha}}' = {-1\over {\cal H}_{\bf n,n}^{\alpha\alpha} }
         \left[ \omega^2 w_{\bf n}^{\alpha} + W_{\bf n}^{\alpha} \right].
\end{equation}
For each step of the iteration, the frequency $\omega$ must be evaluated
according to an expectation value of the form of Eq.\ (\ref{freq}).
Also, in the interest of numerical efficiency, we actually 
store the matrix elements of ${\cal M}$ and only the diagonal 
elements of ${\cal H=M}^2$. 
Then it is useful to realize that an equivalent expression for 
$W_{\bf n}$ is in terms of the entire ${\cal H}$ operation with 
the diagonal part subtracted out:
\begin{equation}\label{useful}
W_{\bf n}^{\alpha} = \left[ \left({\cal H}\Psi\right)_{\bf n}^{\alpha}
                - {\cal H}_{\bf n,n}^{\alpha\alpha} w_{\bf n}^{\alpha} \right]
\end{equation}
This means Eq.\ (\ref{RawIter}) can be written as
\begin{equation}\label{NewIter}
{w_{\bf n}^{\alpha}}' = w_{\bf n}^{\alpha} -
     {1\over {\cal H}_{\bf n,n}^{\alpha\alpha} }
     \left[ 
             \omega^2 w_{\bf n}^{\alpha} 
           + \left({\cal H}\Psi\right)_{\bf n}^{\alpha} 
    \right].
\end{equation}
The stability of this process is further enhanced if we mix a 
fraction $f<1$ of the updated configuration $\Psi'$ 
[Eq.\ (\ref{NewIter})] with a fraction $(1-f)<1$ of the 
old configuration $\Psi$.
This then finally defines the iteration procedure:
\begin{equation}\label{Iter}
{w_{\bf n}^{\alpha}}' = w_{\bf n}^{\alpha} -
    {f \over {\cal H}_{\bf n,n}^{\alpha\alpha} }
     \left[ 
             \omega^2 w_{\bf n}^{\alpha} 
           + \left({\cal H}\Psi\right)_{\bf n}^{\alpha} 
    \right].
\end{equation}
The equation is to be iterated, together with Eq.\ (\ref{freq}) for the
frequency and the usual normalization, Eq.\ (\ref{HermOver}), 
until a self-consistent convergence is achieved.
We should note also that the application of Eq.\ (\ref{Iter})
can be performed either {\em synchronously} or {\em asynchronously}.
In a synchronous step, the old values $w_{\bf n}^{\alpha}$ are 
updated into new ones ${w_{\bf n}^{\alpha}}'$ simultaneously, or
in parallel.
In  an asynchronous step, which is the usual Gauss-Seidel scheme, 
once a site is updated into the new value ${w_{\bf n}^{\alpha}}'$,
that new value is recycled immediately into the right hand side
of Eq.\ (\ref{freq}), and this new value then modifies the
results for its neighboring sites.
Either way, the value of $\omega^2$ is corrected only after the
entire lattice has been updated.
Both ways work well, although the asynchronous step generally
produces faster convergence.
Just as in the time evolution scheme, a random initial configuration 
of $\Psi$ is used for these iterations.
In our actual calculations, we have used fractions $f$ from $0.6$ 
to as high as $1.9$ (an over-relaxation). 
Values of $f$ below $1.0$ provide
reasonably fast convergence with reliable stability.  
Fractions $f$ that are closer to zero result in slower convergence, 
while fractions closer to one give faster convergence at the expense
of greater instability, where the instability causes convergence
to the highest frequency eigenmode rather than the desired lowest
frequency mode.

The above procedure will tend to the lowest frequency mode.
As for the time evolution scheme, to get higher modes it is only 
necessary to combine the Gauss-Seidel iteration with the 
Gram-Schmidt orthogonalization procedure described in Sec.\ 
\ref{GramSchmidt}.

\subsection{Boundary Conditions}
To completely define the eigenvalue problem, boundary conditions
must be specified.  
The simplest choices are Dirichlet and free boundary conditions.
For Dirichlet boundary conditions, we imagine placing fictitious 
spin vectors ${\bf S}_{\bf n_b}$, fixed to the static vortex directions, 
at sites ${\bf n_b}$ just outside the system.
The sites just inside the system interact with these, leading to
on-site terms in the matrix $M$ due to the boundary, 
see Eq.\ (\ref{elements}b),
\begin{eqnarray}
\label{Dirichlet}
M_{\bf n n}^{\tilde{x}\tilde{y}}  &=& -M_{\bf n n}^{\tilde{y}\tilde{x}} 
\nonumber \\
&=& JS \sum_{\bf n_b} \left[ 
  p_{\bf n}^0 p_{\bf n_b}^0 \cos(\phi_{\bf n}^0-\phi_{\bf n_b}^0)
  +  \lambda m_{\bf n}^0 m_{\bf n_b}^0
  \right].
\end{eqnarray}
These are the only effects of the boundary.

For free boundary conditions, the system is open at the boundary,
and no fictitious spins are needed.
Therefore, the above terms of Eq.\ (\ref{Dirichlet}) are absent.
This is the only difference in the matrix between Dirichlet and 
free boundary conditions.
%

However, in the absence of an in-plane magnetic field,
free boundary conditions lead to a zero frequency mode,
due to the in-plane rotational invariance of the model. 
This mode is unusual in that it cannot be made orthogonal to the other
modes by the overlap integrals discussed above.
Therefore, it needs to be removed from the evolving mode solution
by a different tactic.

Consider its structure for $\lambda<\lambda_c$, i.e., in-plane
vortices, where $m_{\bf n}^0=0$,  $p_{\bf n}^0=1$ on all sites. 
Eqns.\ (\ref{w1w2}) become fairly simple, 
\begin{mathletters}
\begin{equation}
JS \sum_{\bf m=n,n'}  
\cos(\phi_{\bf n}^0 - \phi_{\bf m}^0) 
\left( w_{\bf n}^{\it 2}-w_{\bf m}^{\it 2} \right) 
= i \omega_k w_{\bf n}^{\it 1},
\end{equation}
\begin{equation}
JS \sum_{\bf m=n,n'} \left\{ 
\cos(\phi_{\bf n}^0 - \phi_{\bf m}^0) 
w_{\bf n}^{\it 1}- \lambda w_{\bf m}^{\it 1} \right\}
= i \omega_k w_{\bf n}^{\it 2}.
\end{equation}
\end{mathletters}
For the zero frequency mode, there is the solution,
\begin{equation}
\psi_{0, \bf n} = 
\left(
\begin{array}{c} w_{k, \bf n}^{\it 1} \\  w_{k, \bf n}^{\it 2} \end{array} 
\right)
= 
\left(
\begin{array}{c}  0 \\  \varphi_{0} \end{array} 
\right),
\end{equation}
where $\varphi_{0}$ is an arbitrary constant.
Obviously this mode is not normalizable by (\ref{overlap}), 
and represents the fact that a uniform in-plane rotation costs 
no energy. 
%

For $\lambda>\lambda_c$, i.e., out-of-plane vortices,
there is no great simplification of (\ref{w1w2}), however, 
a short calculation shows that the zero frequency mode is
\begin{equation}
\psi_{0, \bf n} =
\left(
\begin{array}{c} w_{k, \bf n}^{\it 1} \\  w_{k, \bf n}^{\it 2} \end{array} 
\right)
= 
\left(
\begin{array}{c}  0 \\  \varphi_{0} \cos\theta_{\bf n}^0 \end{array} 
\right),
\end{equation}
where $\varphi_{0}$ again is an arbitrary constant, and 
$\theta_{\bf n}^0$ is the static out-of-plane structure of the
vortex.
This wavefunction is slightly diminished in amplitude near
the vortex core, where the spins point out of the easy plane.
One can think of the $w^{\it 2}$ component as a ``coordinate''
($w_{\bf n}^{\it 2}=\varphi_{\bf n}\cos\theta_{\bf n}^{0} \approx$ 
in-plane angle), which has the specified value, and $w^{\it 1}$
as the corresponding conjugate momentum (part of $S^{z}$ due
to the spinwave), which has been set to zero. 

In the evolution to find other modes, any component of the
current solution proportional to $\psi_{0}$ must be removed.
For $\lambda<\lambda_c$, it is fairly obvious that we only need
to enforce the constraints, $\langle w_{\bf n}^{\it 1} \rangle =0$,
$\langle w_{\bf n}^{\it 2} \rangle =0$, where $\langle \quad \rangle$
indicates the spatial average.
These are essentially the conditions that the total conjugate
momentum is zero, and that modes have no uniform in-plane drift
component.

For $\lambda>\lambda_c$, physically, the higher modes must
also have a total conjugate momentum equal to zero, and,
clearly, must satisfy $\langle \phi_{\bf n} \rangle = 0$.
This leads to the general constraints we applied for free 
boundary conditions,
\begin{mathletters}
\label{constraint}
\begin{equation}
\langle \delta S_{\bf n}^{z} \rangle = 
\langle w_{\bf n}^{\it 1} \cos\theta_{\bf n}^{0} \rangle = 0,
\end{equation}
\begin{equation}
\langle \varphi_{\bf n} \rangle = 
\langle w_{\bf n}^{\it 2} / \cos\theta_{\bf n}^{0} \rangle = 0,
\end{equation}
\end{mathletters}
where we used 
$S_{\bf n}^{z} = S_{\bf n}^{0 \; z} + \delta S_{\bf n}^{z} =
\sin\theta_{\bf n}^{0} + w_{\bf n}^{\it 1} \cos\theta_{\bf n}^{0}$
to define the dynamical conjugate momentum, $\delta S_{\bf n}^{z}$.
%

\subsection{Memory and CPU Considerations}
These relaxation methods use considerably less memory than the 
diagonalization scheme, provided that we are satisfied to obtain only
a few of the lowest frequency modes, which is the case here.
In order to obtain good stability for systems up to radius $R=100a$,
it is necessary to use double precision. 

If we want to calculate the lowest $K$ modes of 
a system $N$ sites, then we need to save $4N$ complex numbers for each 
state to be calculated (creation and annihilation operators), or 
$8NK$ doubles for all states.
The matrix elements of ${\cal M}$ are real and occupy 
$(4 \times 4 + 2)\times N = 18N$ memory locations, where the factor
accounts for not storing the elements that are zero.
For either type of relaxation scheme, it is convenient to have two 
complex arrays each with $2N$ elements to hold ${\cal M}\Psi$ and 
${\cal M}^2 \Psi$ ($8N$ doubles).
For the Runge-Kutta time evolution, another two complex arrays of size 
$2N$ are needed. 
For the Gauss-Seidel iteration,  on the other hand, the real terms
 ${f \over {\cal H}_{\bf n,n}^{\alpha\alpha} }$ are saved, requiring
$N$ memory locations.
For the asynchronous Gauss-Seidel, cpu time is reduced by
saving ${\cal H}_{\bf n,n''}$, which requires $4 \times 12 \times N$
doubles (there is a total of 12 nearest and next-nearest neighbors
on the square lattice).
Thus the total program size required for a calculation at a single
value of anisotropy $\lambda$ scales as $(34+8K)N$ for Runge-Kutta and
as $(27+8K)N$ (synchronous) or as $(75+8K)N$ (asynchronous) for 
Gauss-Seidel.

For a circular system of radius $R=20a$, with $N=1264$, and choosing 
$K=15$, these come to 1.52 MB for Runge-Kutta and 1.45 MB or 1.93 MB for 
Gauss-Seidel, considerably smaller than the approximately 100 MB 
necessary by diagonalization.
The required memory increases if we want to track the changes 
in the modes with changing anisotropy, which requires storage of the
set of modes at {\em two} different values of $\lambda$.
The primary drawback of these methods is that the cpu time may 
be quite long, because the solution method is essentially diffusive. 
We find that the cpu time increases as $R^4$, where one factor of
$R^2$ is the increase in system area, and the other factor of $R^2$ 
is due to the diffusion time increasing, $t_{\rm diffuse}\propto  R^2$.
Often we are interested to calculate for a set of 20 to 40 values 
of $\lambda$ between zero and one; the large cpu time has prevented us 
from going beyond system radius $R=100a$, where each mode may take on 
the order of a day on typical 100 MHz processors.

It is also clear that the rate at which each mode converges diminishes
with increasing system size, because the frequency differences between
modes decreases as $1/R$. 
Furthermore, it is especially difficult to converge to any mode that is
barely split from the next higher mode.
However, the doubly degenerate modes present in this magnetic spinwave 
problem do not cause difficulties. 
Any arbitrary linear 
combination of such a pair will solve Eq.\ (\ref{Schrod}), and
then the Gram-Schmidt process during the relaxation to the next
higher mode will always produce another equal frequency mode,
orthogonal to the first, spanning the degenerate subspace.

\section{Relation to Continuum Theories}
\label{Continuum}
For a full theoretical description of vortex-spinwave interactions
we need to mention how to compare our results for the wavefunctions
with the spin fields determined in continuum limit theories.
This is straightforward for the FM model, for which we will only make
a few comments.
For the AFM model, however, we are usually analyzing the 
{\em staggered magnetization} $\vec\ell$.
Below we describe how $\vec\ell$ is related to the $w_{\bf n}^{\alpha}$
fields.

\subsection{FM Continuum Coordinates}
\label{FMCont}
For the calculations on a lattice described so far, it was most convenient
to use the Cartesian spin components.
To make some comparison to theory we used {\em planar} spherical 
coordinates [Eq.\ (\ref{spherical})], with $\theta_{\rm planar}$ measured from
the $xy$-plane, such that $\theta_{\rm planar}$ is conveniently
zero in the ground state.
Of course, in much of the literature \cite{} it is common to use
{\em polar} spherical coordinates, with $\theta_{\rm polar}$ measured from the
positive $z$-axis.
Clearly, these are related by $\theta_{\rm polar} = \pi/2 - \theta_{\rm planar}$.
Then the relation that replaces (\ref{varangles}), between our local 
Cartesian components and the angular fluctuations, using polar spherical 
coordinates, is
\begin{equation}
\label{polarvarangles} 
S_{\bf n}^{\tilde{x}} = S \varphi_{\bf n} \sin \theta_{\bf n}^{0}, ~~~
S_{\bf n}^{\tilde{y}} = - S \vartheta_{\bf n}.
\end{equation}
The reason we mention this is that the quantity 
$\mu = \varphi({\bf r}) \sin \theta^{0}({\bf r})$ appears naturally
in the continuum theory for FM vortex-spinwave scattering,\cite{Ivanov+98}
and it is equivalent as well, via Eq.\ (\ref{SxSy}), to our 
$w_{\bf n}^{\it 2}$ field.
Similarly, our $w_{\bf n}^{\it 1}$ field represents the $\vartheta$-fluctuations
discussed in continuum theory.

\subsection{AFM Continuum Coordinates}
\label{AFMCont}
For the AFM scattering problem, there have been two primary ways to
approach the continuum description, where we consider only lattices
composed of two sublattices.

For example, in Refs.\ \onlinecite{Pereira+96,Ivanov+96}, the theory 
is developed directly in terms of $\vec\ell$, defined by the difference 
of spins on the two sublattices,
\begin{equation}
\label{ellvec}
\vec\ell = \frac{ {\bf S}_{\bn} - {\bf S}_{\bn^{\prime}} }{2S}
= \tilde{z}_{\bn} 
+ \frac{S_{\bn}^{\tilde{x}} +S_{\bn^{\prime}}^{\tilde{x}}}{2S} \tilde{x}_{\bn}
+ \frac{S_{\bn}^{\tilde{y}} -S_{\bn^{\prime}}^{\tilde{y}}}{2S} \tilde{y}_{\bn},
\end{equation}
where the local axes are related by 
$\tilde{z}_{\bn^{\prime}} = -\tilde{z}_{\bn}$,
$\tilde{x}_{\bn^{\prime}} = -\tilde{x}_{\bn}$,
$\tilde{y}_{\bn^{\prime}} = \tilde{y}_{\bn}$, and $\bn, \bn^{\prime}$
are neighboring sites, on the two different sublattices.
Assuming $\vec\ell$ has perturbations and is described by planar 
spherical angles
$\phi_{\ell} = \phi_{\ell}^{0}+\varphi_{\ell}$, 
$\theta_{\ell} = \theta_{\ell}^{0}+\vartheta_{\ell}$,
and using Eq.\ (\ref{SxSy}), we then obtain the equivalences,
\begin{mathletters}
\label{equivs}
\begin{equation}
\ell^{\tilde{x}}
=\frac{S_{\bn}^{\tilde{x}}+S_{{\bn}^{\prime}}^{\tilde{x}}}{2S}
=\frac{w_{\bn}^{\it 2}+w_{{\bn}^{\prime}}^{\it 2}}{2S} 
=\varphi_{\ell}\cos\theta_{\ell}^{0} ,
\end{equation}
\begin{equation}
\ell^{\tilde{y}}
=\frac{S_{\bn}^{\tilde{y}}-S_{{\bn}^{\prime}}^{\tilde{y}}}{2S}
=\frac{-w_{\bn}^{\it 1}+w_{{\bn}^{\prime}}^{\it 1}}{2S} 
=\vartheta_{\ell} .
\end{equation}
\end{mathletters}
Here the first equation gives the in-plane fluctuations while the
second gives the out-of-plane tilting fluctuations.
The mapping to the $w_{\bn}^{\alpha}$ fields is correct up to an
arbitrary phase. 
If polar spherical coordinates are used, then change 
$\cos\theta_{\ell}^{0} \rightarrow \sin\theta_{\ell}^{0}$ 
and $\vartheta_{\ell} \rightarrow -\vartheta_{\ell}$.
Once again, it is found that $\ell^{\tilde{x}}$ is equivalent to the
variable $\mu_{\ell}$ introduced by Ivanov {\it et al,}\cite{Ivanov+96}
that appears together with $\vartheta_{\ell}$ in symmetrical 
Schrodinger-like equations.

Alternatively, in other vortex-magnon scattering articles,\cite{PP96}
the spins on the two sublattices have been represented in polar
spherical coordinates introduced by Mikeska,\cite{Mikeska80}
using four angular fields, as 
\begin{equation} \label{Mikeska}
{\bf S} = \pm S \left(
  \begin{array}{c}      \sin(\Theta\pm\theta) \cos(\Phi\pm\phi) \\
                        \sin(\Theta\pm\theta) \sin(\Phi\pm\phi) \\
                        \cos(\Theta\pm\theta) 
  \end{array}           \right) ,
\end{equation}
where for simplicity we suppressed the site indices, and upper/lower
signs refer to the two sublattices. 
The ``small'' angles $\theta$ and $\phi$ are slave to the ``large''
angles $\Theta$ and $\Phi$, via $\theta=-\dot{\Phi}\sin\Theta/8JS$,
and $\phi=\dot{\Theta}/(8JS\sin\Theta)$, showing that both are
truly small for low-frequency modes. 
Then assuming that the large angles have small perturbations
away from a vortex configuration $\Theta^{0},\Phi^{0}$,
$\Theta=\Theta^{0}+\vartheta$, $\Phi=\Phi^{0}+\varphi$,
a short calculation show that the staggered magnetization will
have components,
$\ell^{\tilde{x}} = \varphi \sin\Theta^{0}$,
$\ell^{\tilde{y}} = - \vartheta$.
We see that these are equivalent to the perturbation variables 
in Eq.\ (\ref{equivs}), once the change from polar to planar
coordinates is taken into account.
%

\section{Application: Vortex-Spinwave Spectra and S-Matrix}
\label{Spectra}
The methods described above were tested by comparing their results
with numerical diagonalization results\cite{WysinVolkel96} for small 
systems (up to radius $R\approx 20a$).
At these small system sizes, it was even possible to get very
precise results (more than 6 digits for eigenvalues) for as many as 
the lowest 50 modes.   
Typically, for $\lambda\ne 0$, the mode frequencies for the AFM model 
are higher than those for the FM model, which means that the calculations
tend to converge faster for the AFM model.

As an example of the utility of these methods, 
we consider the spinwave spectrum (lowest 30 modes)
for a single vortex in a circular system, with Dirichlet 
boundary conditions, and use the results to calculate the
scattering S-matrix.
We used a square lattice of spins.
For most of the data presented, the synchronous self-consistent Gauss-Seidel 
scheme with fraction $f=0.7$ was used.  
The FM model for $\lambda$ very close to the critical value is
an exception; there we used the asynchronous sheme, with $f=0.9$,
which has greater stability at the expense of greater CPU time.

Typical mode spectra for some small systems were given in 
Ref.\ \onlinecite{WysinVolkel96}, as functions of the anisotropy
parameter $\lambda$.
In the present work we can consider system radii as large
as $R=100a$, however, the CPU time to get the complete
dependence versus $\lambda$ at such a large radius is
prohibitive, and for most calculations we used $R\le 50a$.
We concentrate on the spectra at a few isolated
values of $\lambda<\lambda_c$, corresponding to in-plane
vortices, and focus more on the variations
with system radius $R$.

\subsection{FM/AFM In-plane vortex S-Matrix}
Due to azimuthal symmetry, the modes generally are classified
according to azimuthal quantum number $m$, corresponding to an
expected $e^{i m \chi}$ dependence, where a point in the system
has polar coordinates $(r,\chi)$, measured from the center. 
Similarly, a principal quantum number $n$ can be defined, that
is the number of nodes in a radial direction in the wavefunction.
There are only minor differences in calculating a scattering
amplitude $\rho_m(k)$ [defined below, see Eq.\ (\ref{rho-def})]
for the FM model compared to the AFM model.
For FM {\em out-of-plane} vortices, the rotational
symmetry around the $S^z$ axis is broken by the nonzero
topological charge and nonzero $S^z$ spin components, 
and generally, $m$ and $-m$ modes are not degenerate.
For the in-plane case we consider here, however, the
topological charge is zero, rotational symmetry holds,
and $\pm m$ modes are degenerate.
In the continuum limit, the zero-field AFM model has complete rotational
and inversion symmetry around the $S^z$ axis, with no topological
charge, so that the modes $\pm m$ are degenerate at any value of $\lambda$.
For both FM and AFM in-plane vortices, the lack of $S^z$ spin
components insures degeneracy of $\pm m$ modes.
The degeneracy is partially broken on a lattice;\cite{Ivanov+98}
for even $m$, the eigenfunctions generally are mixtures of $+m$ and $-m$ 
components.
%

We can analyze the scattering S-matrix both for FM and AFM models as follows.
In the continuum description,\cite{Ivanov+96} the lowest modes,
which belong to the acoustic spinwave branch, have the asymptotic form,
\begin{mathletters}
\label{rho-def}
\begin{equation}
w^{\it 1}(r,\chi) \approx 0,  
\end{equation}
\begin{equation}
\label{asymp}
w^{\it 2}(r,\chi) \sim 
\left[ J_{m}(kr) +  \rho_m(k) Y_{m}(kr) \right] e^{im\chi} ,
\end{equation}
\end{mathletters}
where $J_m$ and $Y_m$ are Bessel and Neumann functions, and $k$ 
is the wavevector as determined from the frequency of the mode.
For the AFM model, the free spinwave dispersion relation is
\begin{equation}
\label{dispersion}
\omega_{\bf k} = 4JS \sqrt{ (1 \mp \gamma_{\bf k}) 
(1 \pm \lambda \gamma_{\bf k}) },
\end{equation}
where upper/lower signs are for the acoustic/optical branch, and
\begin{equation}
\gamma_{\bf k} = \frac{1}{2} ( \cos k_x + \cos k_y ).
\end{equation}
For the FM model, we have only an acoustic branch with dispersion
relation
\begin{equation}
\label{FMdisp}
\omega_{\bf k} = 4JS \sqrt{ (1 - \gamma_{\bf k}) 
(1 - \lambda \gamma_{\bf k}) }.
\end{equation}
Note that the AFM acoustic branch dispersion can be obtained from the 
FM dispersion simply by the change $\lambda\rightarrow -\lambda$.
After the relaxation scheme is used to obtain a mode frequency
$\omega$, then we invert the dispersion relation to get the
corresponding wavevector needed in the fitting expression (\ref{asymp}).
For the FM dispersion, assuming ${\bf k}=(k,0)$, we have 
$k=\cos^{-1}(2\gamma_{\bf k}-1)$, where
\begin{equation}
\label{gamma-k}
\gamma_{\bf k} = (2\lambda)^{-1}\left\{(1+\lambda)
-\sqrt{(1+\lambda)^{2}-4\lambda[1-(\omega/4JS)^2] } \right\}.
\end{equation}
The same formula applies to the AFM acoustic branch when the
change $\lambda\rightarrow -\lambda$ is made.
%

In Eq.\ (\ref{rho-def}), the coefficient $\rho_m(k)$ is a measure of 
the scattering; in the absence of scattering it vanishes, and the 
free spinwave is described by the $J_m(kr)$ function alone.
Alternatively, the $w^{\it 2}$ wavefunction can be expressed using the
S-matrix scattering function, $S_m(k)=e^{2i\Delta_m(k)}$,  where
$\Delta_m(k)$ is the phase shift, in terms of incoming and
outgoing waves, as
\begin{equation}
\label{in+out}
w^{\it 2} \sim \sqrt{\frac{1}{ 2\pi k r}} 
\left\{
e^{-i(kr+\phi_m)} +S_m(k) e^{i(kr+\phi_m)}
\right\},
\end{equation}
where the phase angle is $\phi_m=\frac{\pi}{4}(2m+1)$.
In the absence of scattering, $S_m(k)\rightarrow 1$, 
or $\Delta_m(k)\rightarrow 0$, and
the asymptotic form of $J_m(kr)$ results.
Expressions (\ref{rho-def}) and (\ref{in+out}) are equivalent
up to a constant, provided
\begin{equation}
\label{rho-S}
S_m(k) = \frac{1-i \rho_m(k)}{1+i\rho_m(k)}.
\end{equation}
Then the phase shift is related to $\rho_m(k)$
by
\begin{equation}
\label{Phase}
\tan \Delta_m(k) = - \rho_m(k).
\end{equation}
%

The scattering amplitude $\rho_m(k)$ can be found by making
a least squares fitting of a wavefunction to the above 
asymptotic form (\ref{asymp}).
However, in the case $m>0$, when $m$ and $-m$ 
are degenerate, we actually fit to the expression,
\begin{eqnarray}
\label{expression}
w^{\it 2}(r,\chi) & = & 
\left[ a_1 J_{m}(kr) +  b_1 Y_{m}(kr) \right] e^{im\chi} 
\nonumber \\
&+&
\left[ a_2 J_{m}(kr) +  b_2 Y_{m}(kr) \right] e^{-im\chi} ,
\end{eqnarray}
where $a_1$, $a_2$, $b_1$, and $b_2$ are complex fitting constants.
Then, $\rho_m(k)=b_1/a_1$ and $\rho_{-m}(k)=b_2/a_2$ are extracted,
and should be expected to be comparable.
For in-plane vortices, whose static structure has no 
$\lambda$-dependence, we considered the asymptotic region to
be $r>8a$.

\subsection{The Scattering Results: In-Plane Vortices}
We considered in-plane FM and AFM vortices for $\lambda=0.0, 0.5, 0.7$.
The static in-plane vortex spin structure is confined to the
xy plane, resulting in a singular point at the vortex core,
and making continuum theory difficult.
For any $\lambda<\lambda_c$, the static structure of in-plane FM
vortices is the same as the in-plane AFM vortex structure on
one sublattice.
However, the dynamics are different, in particular, the FM
mode frequencies tend to be much lower than those for the AFM,
although the appearances of their wavefunctions are very similar.
The only exception to this is at $\lambda=0$, where the FM and
AFM models have the same mode frequencies.  
Thus it is interesting to compare the FM/AFM scattering properties
and also investigate the anisotropy dependence.

\begin{figure}
\psfig{figure=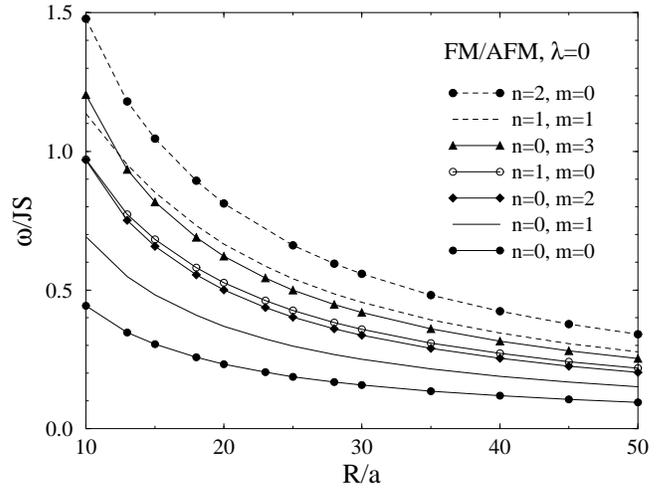,angle=-90.0,width=\pssize pt}
\caption{
\label{modes-R-0}
The frequencies of some of the lowest modes for a single FM or AFM in-plane
vortex, at $\lambda=0.0$, versus system radius $R$. The integers $(n,m)$
label the principal and azimuthal quantum numbers.}
\end{figure}

\begin{figure}
\psfig{figure=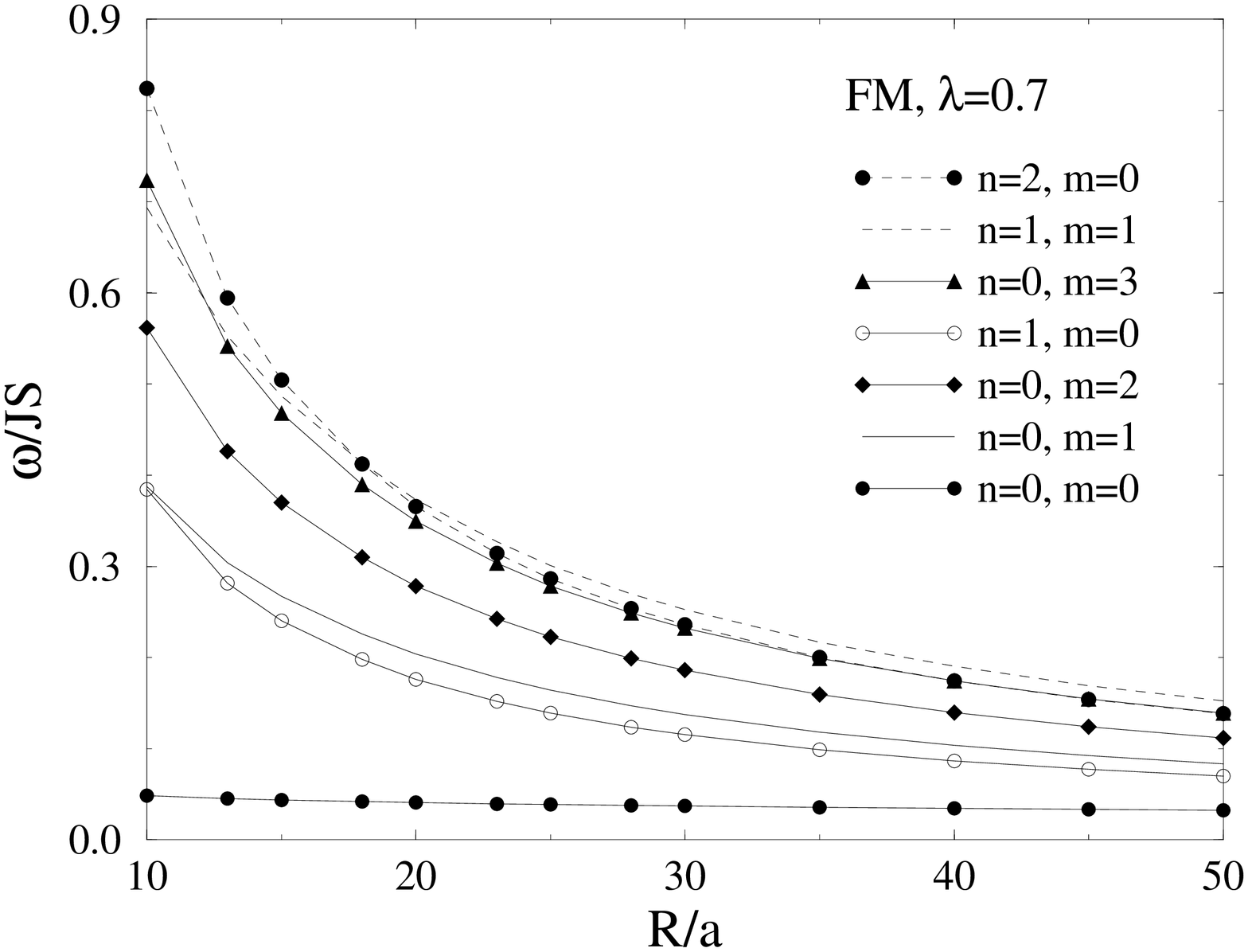,angle=-90.0,width=\pssize pt}
\psfig{figure=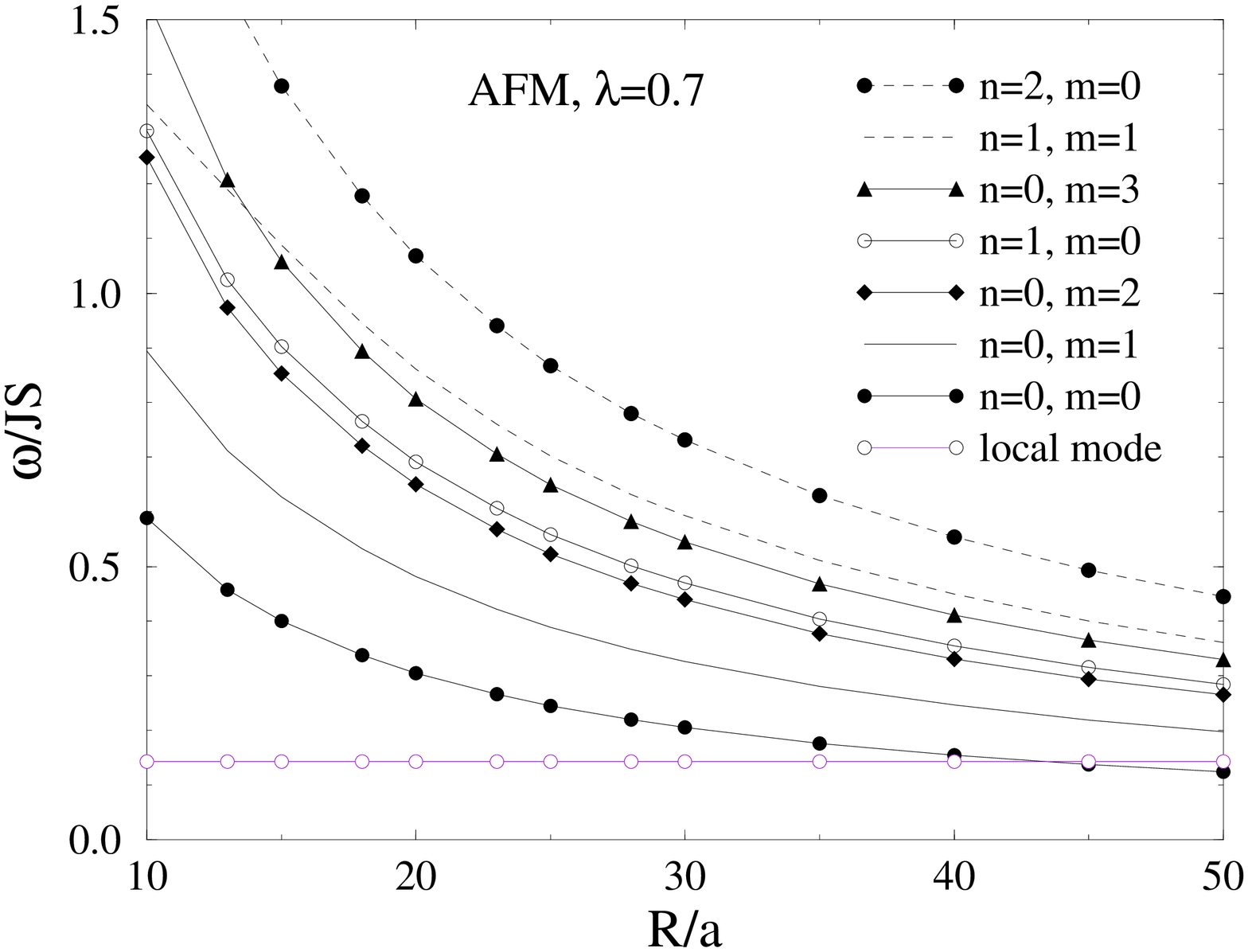,angle=-90.0,width=\pssize pt}
\caption{
\label{modes-R-7}
The frequencies of some of the lowest modes for a single a) FM or b)
AFM in-plane vortex, at $\lambda=0.7 < \lambda_c$, versus system radius $R$.   
}
\end{figure}

Wavefunctions of some of the lowest modes were presented in 
Refs.\ \onlinecite{Ivanov+96,WysinVolkel96}, and more complete 
wavefunction and other data, from the new methods
described here, can be found at 
http://www.phys.ksu.edu/\~\ wysin/vortexmodes/ .
From those diagrams it is easy to identify the principal
and azimuthal quantum numbers $n$ and $m$ for each mode.  
In Fig.\ \ref{modes-R-0} we show the dependence of some of 
the lowest eigenfrequencies on system radius $R$, for
$\lambda=0$ (identical results for FM and AFM here).
Differences appear in these spectra as $\lambda$
approaches the critical value, as seen in Fig.\ \ref{modes-R-7},
where the spectra are shown for $\lambda=0.70$, which is just
below $\lambda_c \approx 0.7034$ .
Additional data for $\lambda=0.5$ are presented at 
the www page listed above.
With the exception of the local mode in the AFM, most modes 
have frequencies diminishing asymptotically as $1/R$, as expected 
in a continuum theory.
The local mode in the AFM model has a frequency independent
of the system size, but dependent on $\lambda$, and can be 
thought of as a state taken out of the optical spinwave branch.  
The lowest mode in the FM model is only quasi-local: its wavefunction
is local in the out-of-plane fluctuations but extended in the
in-plane fluctuations, and its frequency decreases slowly with
increasing system size.

We calculated the spectrum of the lowest 30 modes for different
system sizes $R$, ranging from $R=13a$ to $R=100a$.
Inverting the $\omega_k$ relationship (\ref{FMdisp}) as described 
above [via Eq.\ (\ref{gamma-k})], and by using these different systems 
sizes, we could observe a particular mode labeled by $(n,m)$ at 
different $k$.
In this way we obtained the $k$-dependence of the scattering.
%

\subsubsection{Scattering by FM Vortices}
\begin{figure}
\psfig{figure=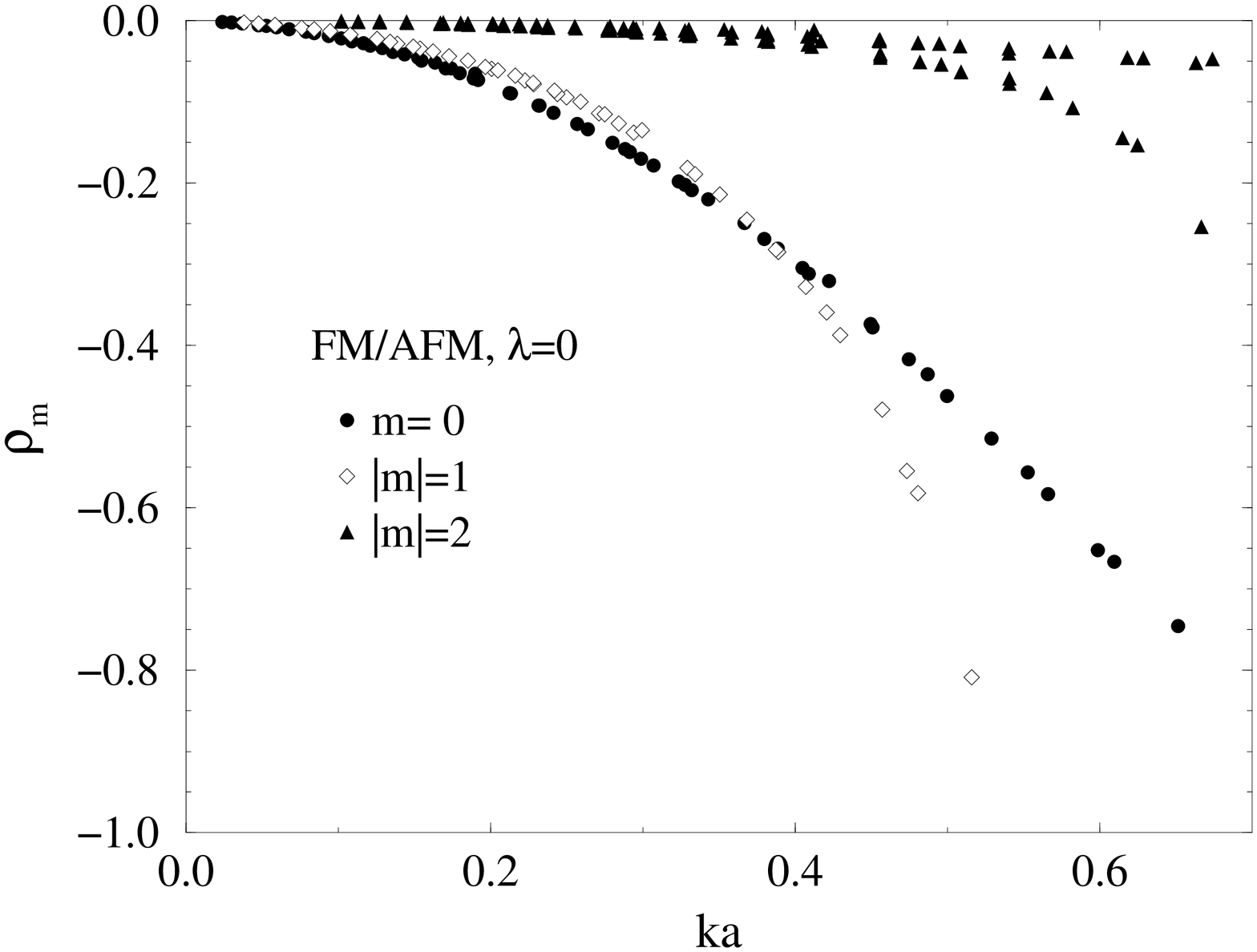,angle=-90.0,width=\pssize pt}
\psfig{figure=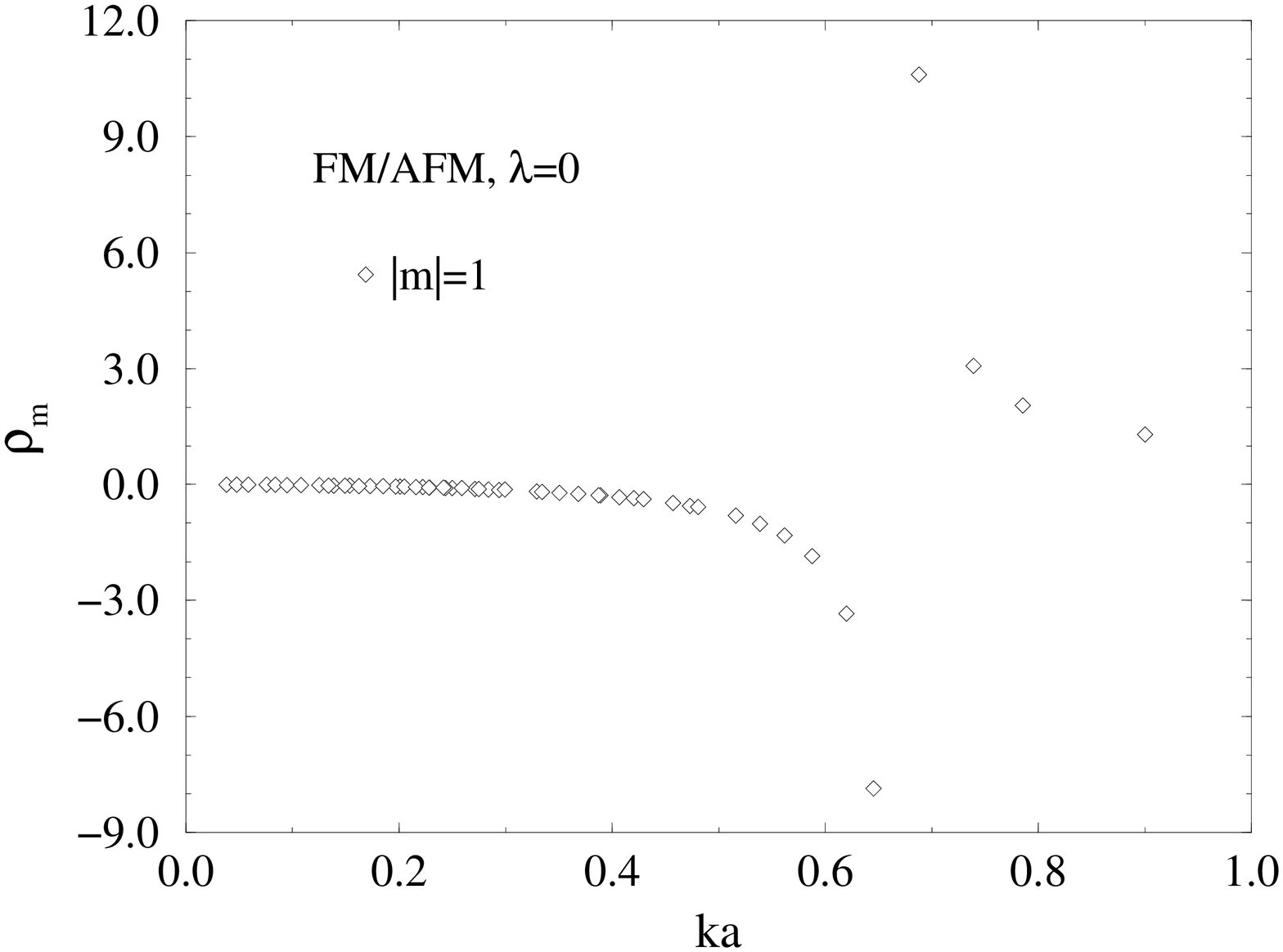,angle=-90.0,width=\pssize pt}
\caption{
\label{rho-FM-0}
The lowest scattering amplitudes $\rho_m(k)$ for a FM or AFM in-plane 
vortex, at $\lambda=0.0$, for a) $m= 0, 1, 2$ and b) expanded
view of the $m=1$ amplitude, exhibiting the singularity.}
\end{figure}

\begin{figure}
\psfig{figure=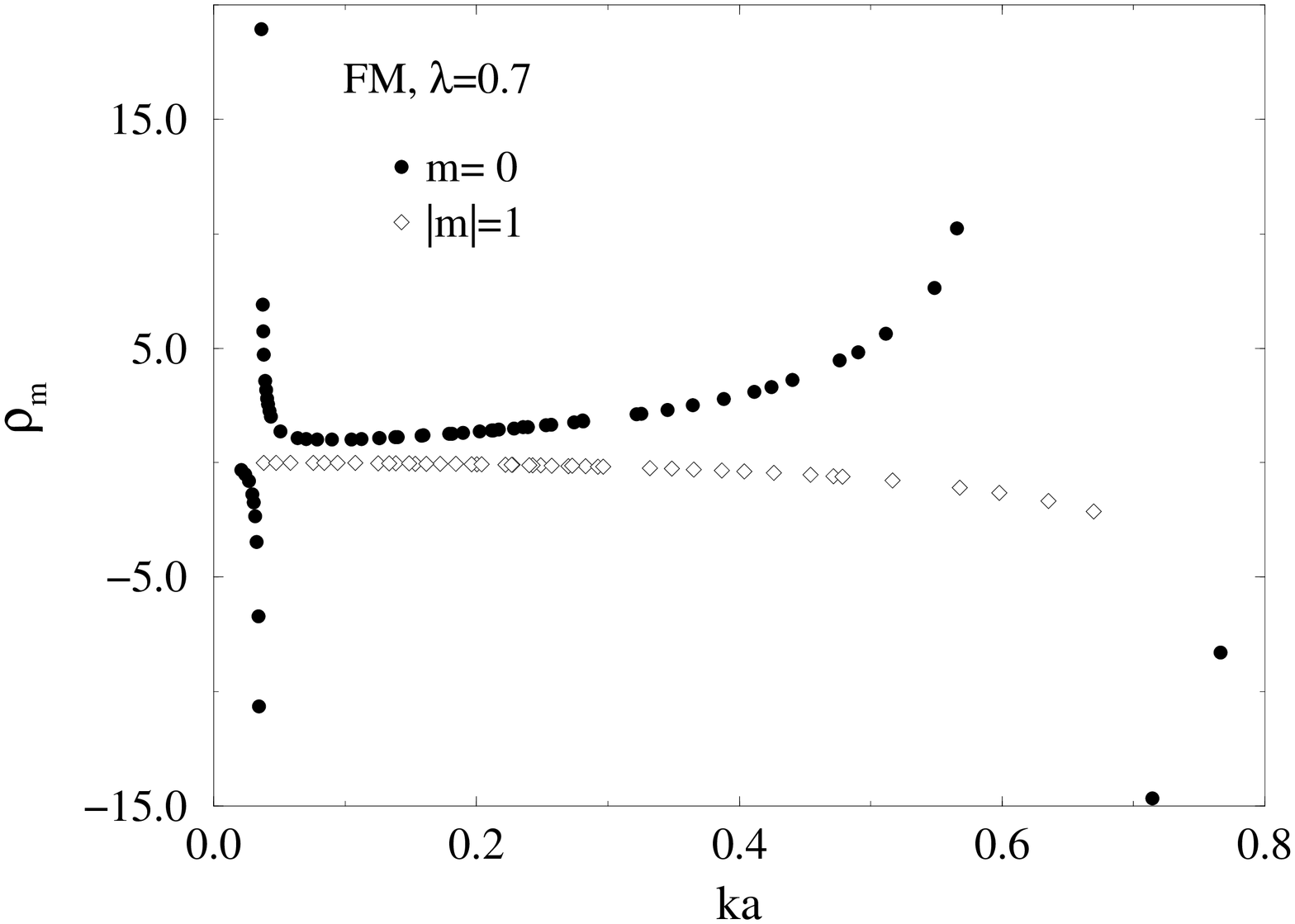,angle=-90.0,width=\pssize pt}
\psfig{figure=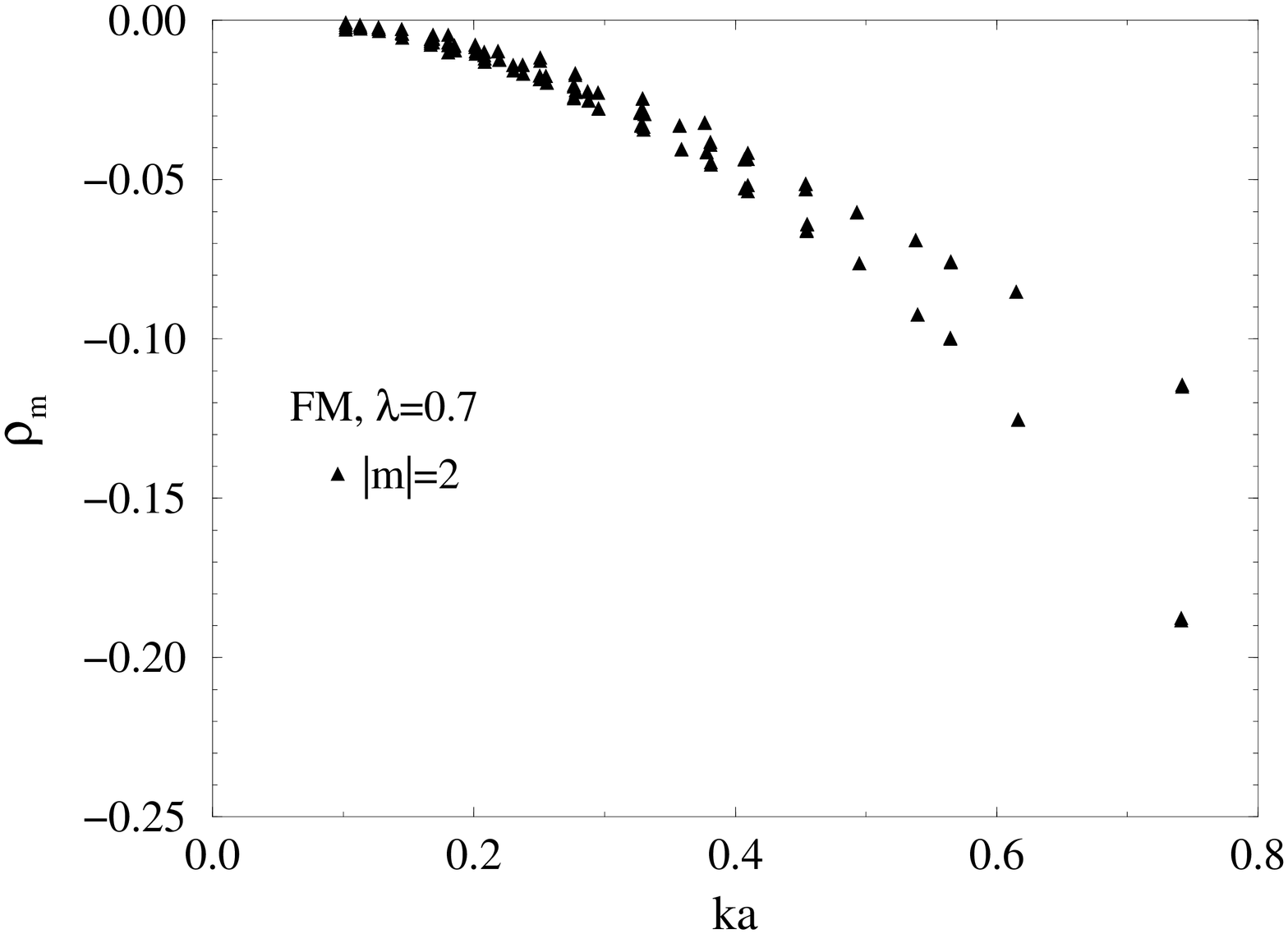,angle=-90.0,width=\pssize pt}
\caption{
\label{rho-FM-7}
The lowest scattering amplitudes $\rho_m(k)$ for a FM
vortex, at $\lambda=0.7 < \lambda_c$, for a) $m=0, 1$ and b) $m=2$.}
\end{figure}

Some results obtained for the scattering of spinwaves by FM vortices
are shown in Fig.\ \ref{rho-FM-0} and Fig.\  \ref{rho-FM-7}, 
where the scattering amplitude $\rho_m(k)$ is displayed for $\lambda=0.0, 0.7$.
The scattering tends to be much weaker as $m$ increases above 2.
As $k \rightarrow 0$, $\rho_m(k)$ tends to zero, a physically
reasonable limit.
An interesting result, especially as $\lambda\rightarrow \lambda_c$,
is that $\rho_m(k)$ has points where it is singular and changes sign.
Indeed, the $m=0$ scattering curve at $\lambda=0.7$ exhibits two
singularities, one near $ka=0.035$ and another near $ka=0.65$. 
Note that the scattering results obtained for $\lambda=0.0, 0.5$,
probe down to $ka=0.024$; down to this point they do not
display the lower singularity.
%

\begin{figure}
\psfig{figure=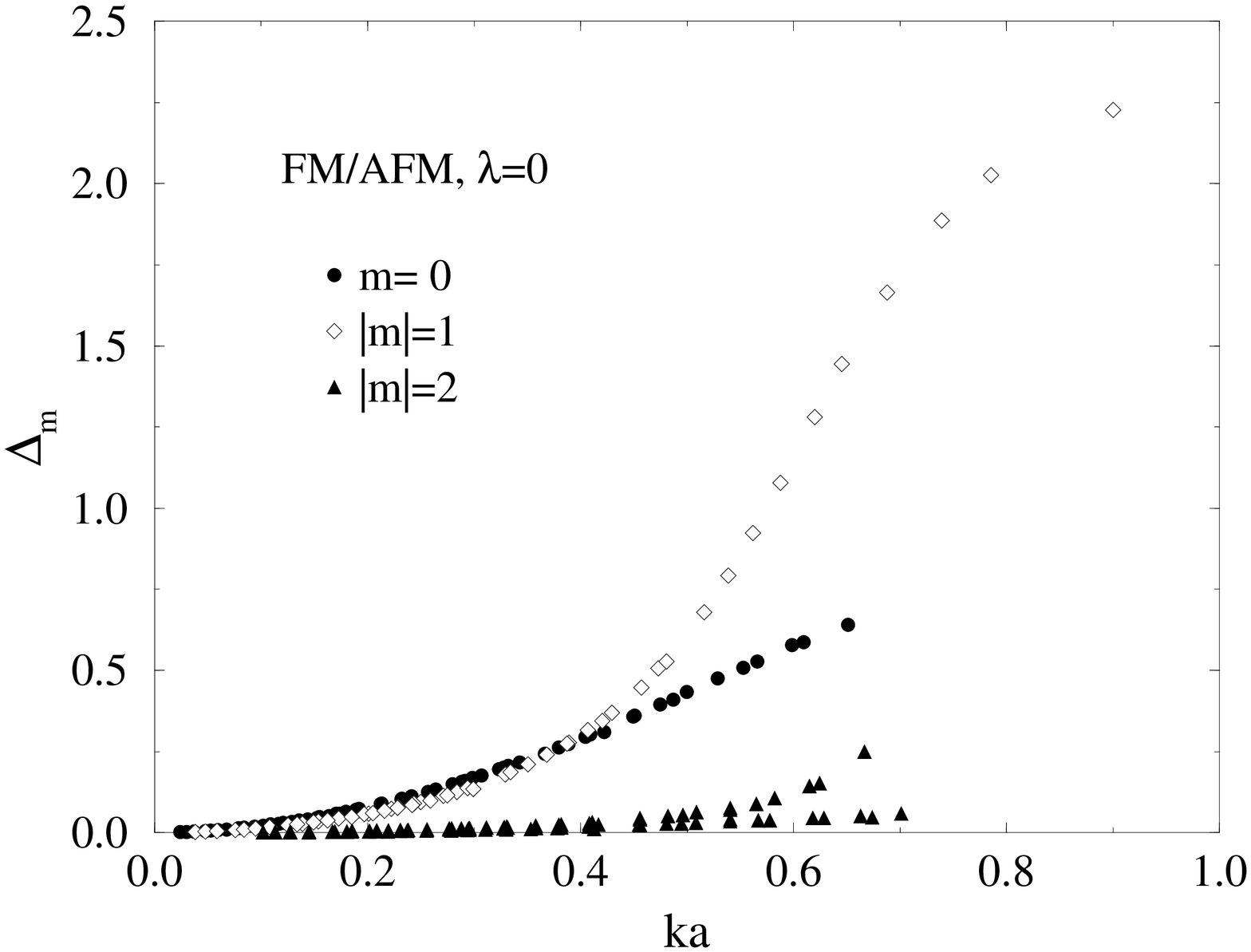,angle=-90.0,width=\pssize pt}
\caption{
\label{Delta-FM-0}
The lowest scattering phase shifts $\Delta_m(k)$ for a FM or AFM in-plane 
vortex, at $\lambda=0.0$, for $m= 0, 1, 2$.}
\end{figure}

\begin{figure}
\psfig{figure=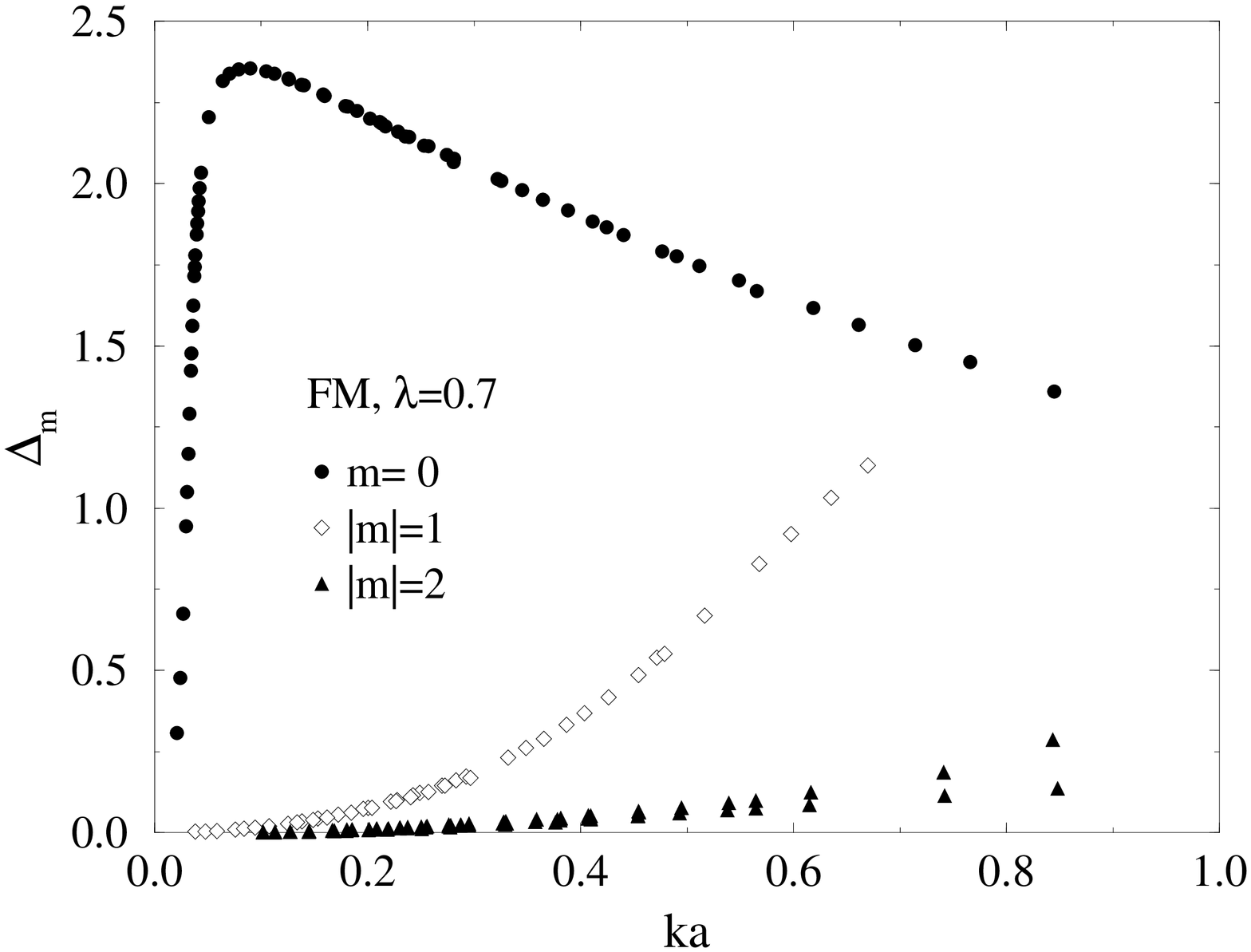,angle=-90.0,width=\pssize pt}
\caption{
\label{Delta-FM-7}
The lowest scattering phase shifts $\Delta_m(k)$ for a FM in-plane 
vortex, at $\lambda=0.7$, for $m= 0, 1, 2$.}
\end{figure}

Using Eq.\ (\ref{Phase}), we see that these singular points in 
$\rho_m(k)$ correspond to the phaseshift $\Delta_m(k)$ passing 
through the value $\pi/2$ near $ka=0.035$ and near $ka=0.65$.
In Figs.\ \ref{Delta-FM-0} and \ref{Delta-FM-7} we show the 
resulting phaseshifts $\Delta_m(k)$ for $\lambda=0.0, 0.7$;  
the phaseshifts are smooth functions of $k$.
It is impressive that the $m=0$ phase shift for $\lambda=0.7$
rapidly surpasses $\pi/2$, reaches a maximum and
then falls back through $\pi/2$. 
Additional data for other parameters is found at the www page.
Note that as long as $\rho_m(k)\ll 1$ there is little
difference between it and the phase shift, except for the
reverse sign.

Costa {\it et al.}\cite{Costa92} calculated the $m=0$ and $m=1$
phase shifts for the FM model, only for $\lambda=0$, in the 
continuum limit via a Born approximation.
Our results here are somewhat in contrast to theirs; 
from Fig.\ \ref{rho-FM-0} and Eq.\ (\ref{Phase}) one sees that we obtain 
the opposite sign, $\Delta_m(k)>0$ for $m=0,1,2$, and over the range, 
$0\le ka < 0.5$, there are no singularities in $\rho_m(k)$. 
Thus these phase shifts do not reach $\pi/2$ in this interval.
For $m=0$, the Costa {\it et al.} result is similar to ours
(except for the opposite sign), whereas for $m=1$ their
result for the phase shift $\Delta_1(k)$ has reached $-\pi$ 
at $ka\approx 0.3$.
Note, however,  that the analytic calculations require the
(somewhat arbitrary) choice of a short distance cutoff on the
integrals appearing in the Born approximation, at a
distance on the order of a lattice constant ($a/2$ was used
in Ref.\ \onlinecite{Costa92}). 
It is clear that a different cutoff could lead to drastically
modified results, as was shown, for example, in a 
calculation\cite{Wysin+98} of the frequencies of the
local mode of AFM in-plane vortices.  
The choice of how to smoothly convert the singular field in
the core of the vortex into a smooth continuum field is not
clear, and always makes it difficult to write out an unambiguous
continuum theory for the in-plane vortices.
Therefore it may not be so fruitful to make any detailed comparison
of the discrete lattice results here to continuum results for
in-plane vortices, unless no cutoff is used.
(This is not a problem for the out-of-plane vortices, which are not
singular at their core.)

\subsubsection{Scattering by AFM Vortices}
\begin{figure}
\psfig{figure=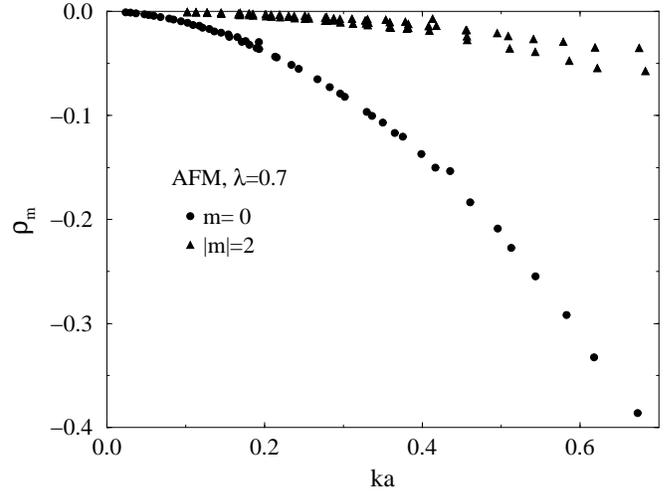,angle=-90.0,width=\pssize pt}
\psfig{figure=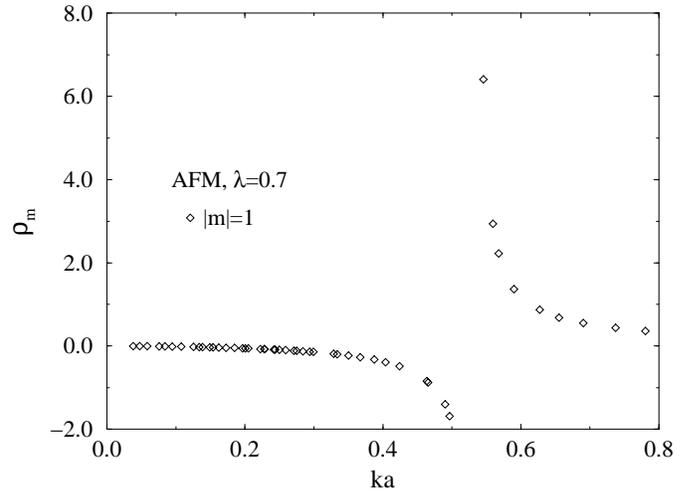,angle=-90.0,width=\pssize pt}
\caption{
\label{rho-AFM-7}
The lowest scattering amplitudes $\rho_m(k)$ for an AFM
vortex, at $\lambda=0.7$, for a) $m=0, 2$ and b) $m=1$, 
showing the singularity between $ka=0.5$ and $ka=0.6$ .} 
\end{figure}

\begin{figure}
\psfig{figure=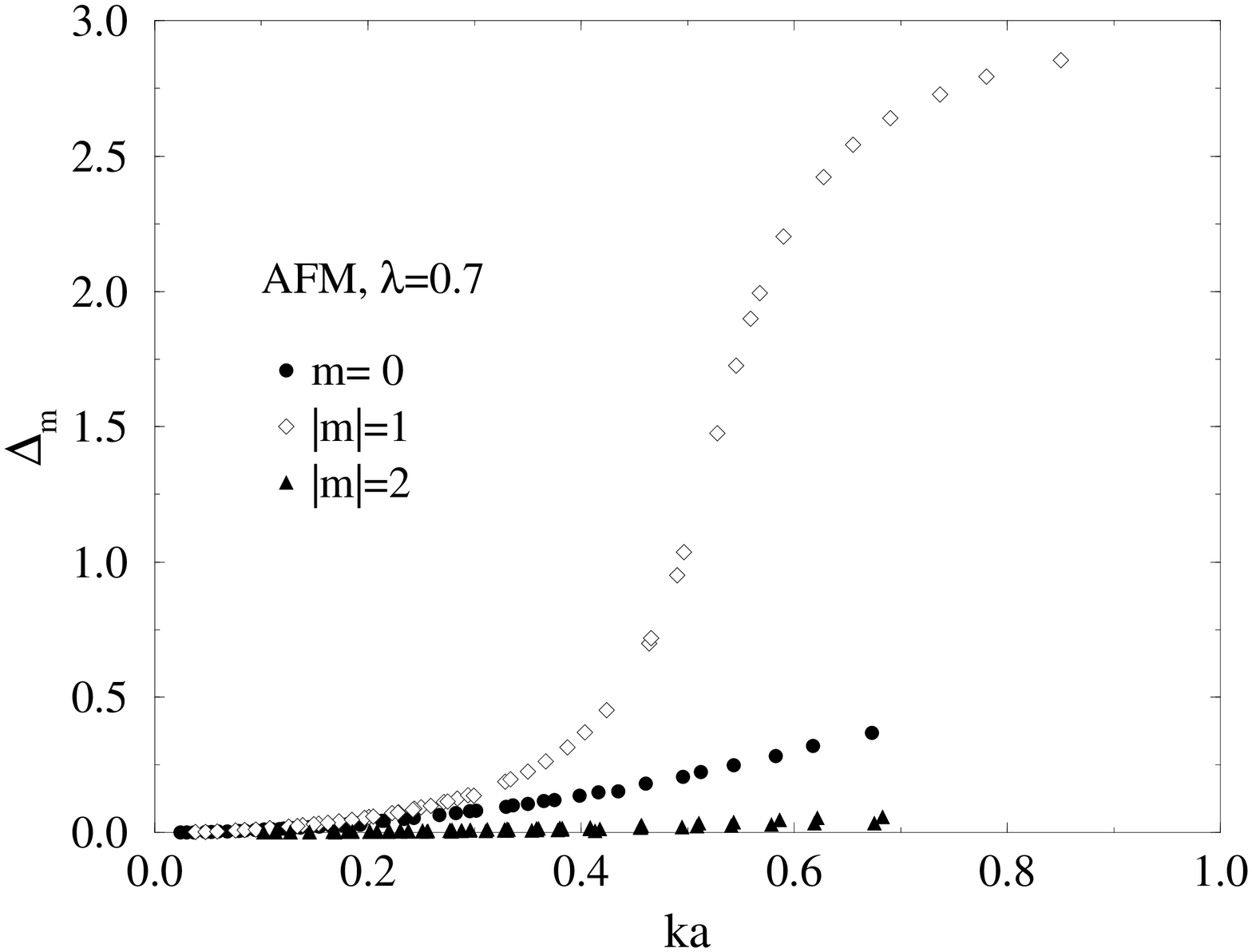,angle=-90.0,width=\pssize pt}
\caption{
\label{Delta-AFM-7}
The lowest scattering phase shifts $\Delta_m(k)$ for an AFM in-plane 
vortex, at $\lambda=0.7$, for $m= 0, 1, 2$.}
\end{figure}

At $\lambda=0$, there is no difference in the scattering for AFM
vortices compared to FM vortices; Figs.\ \ref{rho-FM-0} 
and \ref{Delta-FM-0} apply.
(Only the acoustic AFM branch is being considered; the optical
branch is too high in the spectrum for these relaxation techniques
to be useful.)
Similarly, results for the scattering by AFM vortices at
$\lambda=0.7$ are shown in Fig.\ \ref{rho-AFM-7}. 
Results obtained at $\lambda=0.5$ are very similar to these,
and shown on the web page indicated above.
For the $m=1$ scattering, there is a singular point somewhere
around $ka\approx 0.55$; thus $\Delta_1(k)$ passes through
$\pi/2$ at this point, as seen in Fig.\ \ref{Delta-AFM-7}.  
Although the AFM frequencies are higher than the corresponding frequencies
for the FM model, the data obtained probe to similar low values
of $ka\approx 0.024$, which is essentially determined by the largest
system size used, $R=100a$;  no singularity for $m=0$ was found down 
to this wavevector limit.
Therefore we can say that the $m=0$ scattering by AFM in-plane 
vortices, just below $\lambda_c$, is considerably weaker than 
that for FM in-plane vortices.
On the other hand, the $m=1$ scattering in this case is much stronger
by the AFM in-plane vortices than by the FM in-plane vortices. 

Continuum results for the AFM {\em acoustic} branch spinwave 
scattering by {\em in-plane} vortices are intriguing.
Pereira {\it et al.}\cite{Pereira+96} have calculated the scattering
phase shifts for the {\em optical branch}, which excite out-of-plane
spin fluctuations (our $w^{\it 1}$ field), even without the need for a Born
approximation.  
However, in the leading order continuum theory, the potential due
to the vortex appears {\em only} in the perturbation equation for the
out-of-plane fluctuations (angle $\vartheta_{\ell}$); 
it is absent from the equation for the 
fluctuations of the in-plane components (angle $\varphi_{\ell}$ or
our $w^{\it 2}$ field).
Indeed, the dynamical equation for $\varphi_{\ell}$ is simply
a wave equation for free acoustic spinwaves.
It means that the usual continuum limit predicts no scattering
of acoustic spinwaves (which excite in-plane fluctuations)
by in-plane AFM vortices. 
We might comment, however, that for out-of-plane vortices,\cite{Ivanov+96} 
the potential due to the vortex appears in both the equations,
one for $\vartheta_{\ell}$, the other for the quantity 
$\mu_{\ell}=\varphi_{\ell}\cos\theta_{\ell}^{0}$ [Eq.\ (\ref{equivs})].
So the scattering from in-plane vortices is a special case.

Of course, with our numerical calculations on the lattice
for AFM in-plane vortices, we did obtain some relatively weak 
scattering for $m=0,2$, and rather stronger scattering for $m=1$.
It is possible that this unexpected scattering is caused primarily
by the vortex core, where the static spin directions rotate by
$90^{o}$ for nearest neighbor sites.
This gives large gradients in the core region, which cannot be
described in the usual lowest order continuum limits.
Some theoretical analysis of this kind of discrete effect, which
always occurs around the vortex core, is needed.

\section{Vortex Mass}
\label{Mass}
\begin{figure}
\psfig{figure=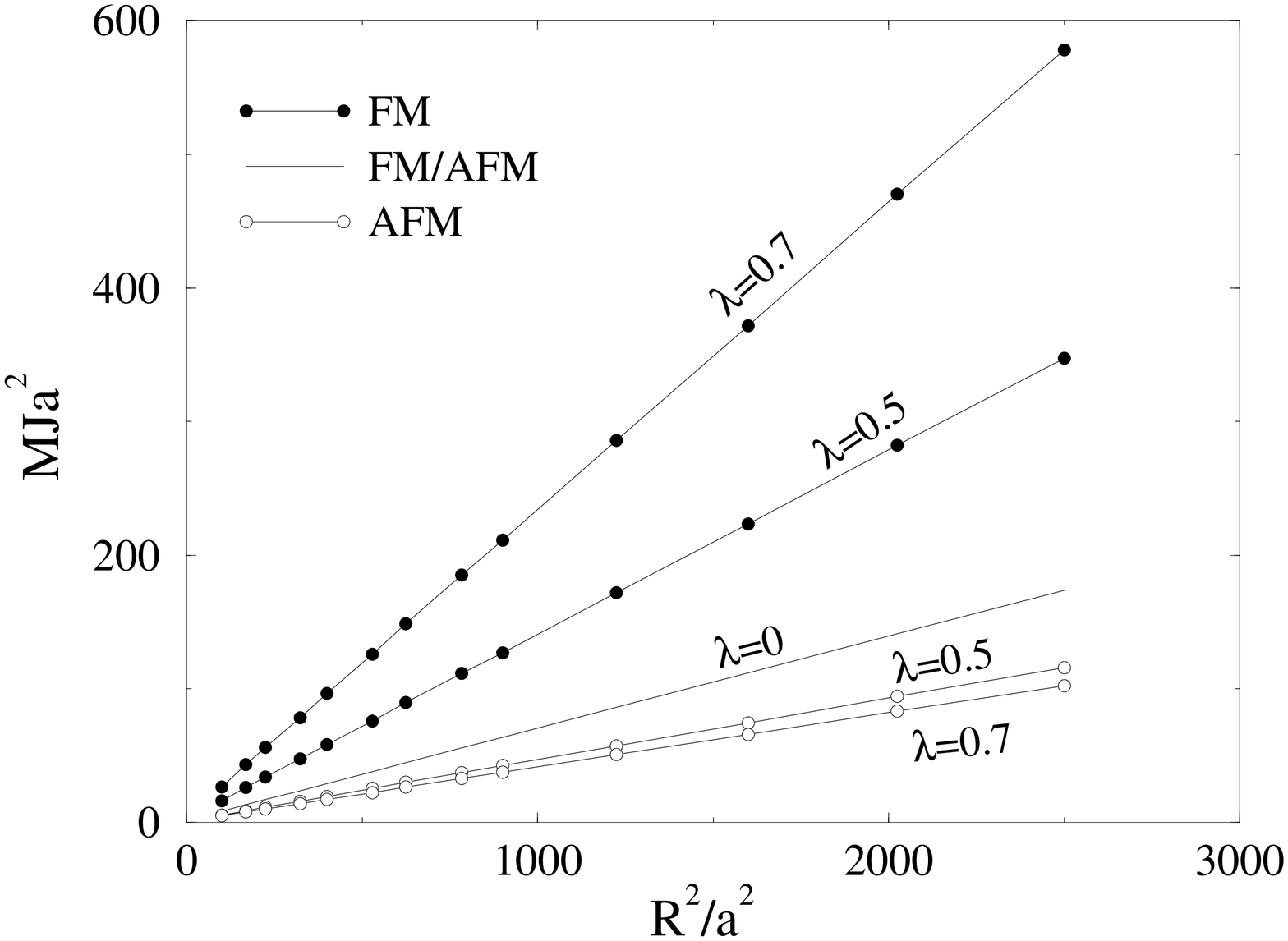,angle=-90.0,width=\pssize pt}
\caption{
\label{masses}
The estimated vortex masses from Eq.\ (\protect\ref{massEq}), vs. 
squared system radius, for indicated anisotropy constants. }
\end{figure}

As another application, information about certain normal modes can
be used to estimate a vortex mass.
The results here apply only to in-plane vortices.
%
The modes with $(n=0, m=1)$, which are doubly degenerate for
in-plane vortices,  are most strongly associated with the 
translational motion of the vortex center.
If only these modes are added to the original static vortex structure,
the resulting configuration produces either a circular or linear
simple harmonic oscillatory motion of the vortex 
center.\cite{WysinVolkel95,Wysin96}
These motions are the same as those predicted from a simple 
Newtonian dynamical equation of motion for the vortex center,
involving a force $\vec{F}$ and effective mass $M$, i.e.,
\begin{equation}
\vec{F} = M\dot{\vec{V}},
\end{equation}
where $\vec{V}$ is the velocity of the vortex center $\vec{X}(t)$.
The translational modes can be considered to be driven by the
force caused by the system boundary and the lattice itself,
which can be evaluated independently of the spinwave spectrum.
In the simplest approximation, we assume a linear restoring
force, $\vec{F}=-K\vec{X}$, where the spring constant $K$ has 
been found by a relaxation scheme applied to a static vortex
in Ref.\ \onlinecite{Wysin96}.
For in-plane vortices, it was found that $K\approx 4 JS^2/a^2$, 
independent of the value of $\lambda$. 
Then, for this simple harmonic motion, the mass is found as
\begin{equation}
\label{massEq}
M=K/ \omega_{0,1}^2  ,
\end{equation}
where $\omega_{0,1}$ is the translation mode frequency.
The results of this calculation are shown in Fig.\ \ref{masses},
for both the FM and AFM models at various anisotropies.
The fact that the translational mode frequencies diminish as
the reciprocal system size is reflected there; the effective
mass increases as the square of the system radius $R$.
The masses are identical for FM and AFM vortices at $\lambda=0$,
where their spinwave spectra are identical.
The FM vortex mass increases with increasing $\lambda$, while the
AFM vortex mass decreases.
Thus we have the result that in some real dynamical sense AFM vortices
will be easier to move than FM vortices.
The fact that the mass depends on the system size should not be
unexpected, because a vortex is not a localized object.
Its spin configuration extends to the limits of the system,
causing its inertia against motion inside the system to depend
on the system size.
The mass defined here really is more a property of the entire system.
%

\section{Conclusions}
\label{Conclude}
We have discussed some relaxation methods for determining
the lower part of the eigenspectrum for magnetic models in the
presence of an individual magnetic nonlinear excitation, such as
a vortex.
The methods are based on the fact that the effective time evolution
of the {\em squared} Hamiltonian is diffusive, and therefore evolution
from an arbitrary intial state will lead to the lowest mode present.
By combining with Gram-Schmidt orthogonalization, a reasonably
large set of the lowest modes, including their eigenfunctions
and eigenfrequencies, can be determined.
Special considerations also were described for applying free
boundary conditions, in contrast to Dirichlet boundary conditions;
for the former the zero-frequency mode must be eliminated from
the evolving solution.

%
As an application of the schemes, the vortex-spinwave scattering
and S-matrix were determined numerically for both FM and AFM in-plane 
vortices on square lattice circular systems.
The results obtained exhibit some interesting singular points in
the scattering amplitude $\rho_m(k)$,
especially as $\lambda$ approaches the critical value from below.
A singularity in the $m=1$ scattering amplitude
by FM or AFM vortices ($\lambda=0.0$)
occurs near $ka\approx 0.65 $.
A singularity in the $m=0$ scattering amplitude for the FM 
at $\lambda=0.7$ occurs at very low wavevector ($ka\approx 0.035$) 
or over long distances, and it may be interesting to ask whether 
it could be described in a continuum theory.
These singular points occur where the phase shift passes 
smoothly through values like $\pm\pi/2$.

At $\lambda=0$, the scattering results for FM and AFM vortices
are identical, as a result of identical mode frequencies.
For the AFM model, although theoretically there should be
no scattering of acoustic spinwaves by a vortex, we did find 
nonzero scattering amplitude, especially for $m=1$, and
even a singular point in this amplitude for $\lambda=0.7$ .
This unexpected strong scattering is probably due to the
large gradients in the spin field around the vortex core, which 
cannot accurately be described in the usual continuum limit
theory.

The lowest vortex translational modes (lowest mode most closely
coupled to motion of the vortex center) were used to estimate a
vortex mass, for in-plane vortices, in a very simple phenomenolgical
theory.
The theory is based on the idea that the vortex in a small system
vibrates in its translational mode in response to the force on it
due to the lattice and system boundary, with a frequency determined
by the effective force constant and the vortex mass.
Using the observed translational mode frequencies, this simple theory 
leads to a vortex mass increasing as the square of the system size, 
for both FM and AFM vortices.
AFM vortices have lower effective mass and thus are expected to be
easier to move.
For the in-plane vortices, FM vortices at $\lambda\simeq \lambda_c$
have the highest mass, while AFM vortices at $\lambda\simeq \lambda_c$
have the lowest mass.

With appropriate modifications to the Hamiltonian, etc., in general, 
the relaxation schemes discussed here would apply to other kinds of 
stable nonuniformly magnetized systems, such as, for example, small 
magnetic particles with a frozen-in magnetic configuration.
Then the methods might be very useful for determining any interesting
response properties of such materials.

\vskip 0.08in
{\sl Acknowledgements.}---The author gratefully acknowledges the
support of  NSF Grant DMR-9412300,  NSF/CNPq International 
Grant INT-9502781, and a FAPEMIG Grant for Visiting Researchers
while at UFMG.  Discussions with M. E. Gouv\^ea, A. S. T. Pires,
and B. A. Ivanov are greatly appreciated.

\ifnum\pageformat=1
  \newpage
\fi

\begin{references}
\bibitem{orthoNote}	The scalar product between modes of different 
frequencies is not the usual scalar product, but rather an overlap between 
right and left eigenvectors, because the underlying spinwave Hamiltonian 
system is not Hermitian.  The appropriate scalar product is given in 
Ref. \protect\onlinecite{WysinVolkel95}.
\bibitem{SignNote}	This transformation appears somewhat more pleasing, 
without minus signs, if we use the matrices ${\cal M}'= i {\cal M}$, 
${\cal A}' = i{\cal A}$, i.e., the relationships are then, 
${\cal A' = A'}^{-1} = {\cal A'}^{\dagger}$, and
 ${\cal A' M' A' = M'}^{\dagger}$.
\end{references}
\end{document}